\documentclass[journal,10pt]{IEEEtran}

\usepackage{cite}

\ifCLASSINFOpdf
\usepackage{graphicx}
  \graphicspath{{../pdf/}{../jpeg/}}
\DeclareGraphicsExtensions{.pdf,.jpeg,.png,.eps}
\else
\fi
\usepackage[cmex10]{amsmath}
\usepackage{amssymb}
\usepackage{mathtools}

\usepackage{siunitx}

\usepackage{algorithm}
\usepackage{algpseudocode}
\usepackage{algorithm}
\newsavebox{\ieeealgbox}

\usepackage[acronym]{glossaries}
\newacronym{mimo}{MIMO}{multiple-input multiple-output}
\newacronym{v2v}{V2V}{vehicle-to-vehicle}
\newacronym{gscm}{GSCM}{geometric-based stochastic channel model}
\newacronym{dmc}{DMC}{diffuse multi-path components}
\newacronym{sp}{SP}{specular paths}
\newacronym{ekf}{EKF}{extended Kalman filter}
\newacronym{sage}{SAGE}{space-alternating generalized expectation-maximization}
\newacronym{io}{IO}{interaction objects}
\newacronym{mle}{MLE}{maximum likelihood estimator}
\newacronym{blue}{BLUE}{best linear unbiased estimator}
\newacronym{tx}{Tx}{transmitter}
\newacronym{rx}{Rx}{receiver}
\newacronym{ddtf}{DDTF}{double directional transfer function}
\newacronym[plural = APDP, firstplural=average power delay profiles (APDP)]{apdp}{APDP}{average power delay profile}
\newacronym{tdoa}{TDoA}{time delay of arrival}
\newacronym{dod}{DoD}{direction of departure}
\newacronym{doa}{DoA}{direction of arrival}
\newacronym{3d}{3D}{3-dimensional}
\newacronym{4d}{4D}{4-dimensional}
\newacronym{cdf}{CDF}{cumulative distribution function}
\newacronym{hrpe}{HRPE}{high resolution parameter estimation}
\newacronym[plural = MPC, firstplural=multipath components (MPC)]{mpc}{MPC}{multipath components}
\newacronym{fim}{FIM}{Fisher Information Matrix}
\newacronym{em}{EM}{Expectation Maximization}
\newacronym{tdm}{TDM}{time-division multiplex}
\newacronym{snr}{SNR}{signal-to-noise ratio}
\newacronym{pdp}{PDP}{power delay profile}
\newacronym{ofdm}{OFDM}{orthogonal frequency division multiplexing}
\newacronym{lo}{LO}{local oscillator}
\newacronym{pps}{PPS}{pulse per second}
\newacronym{pa}{PA}{power amplifier}
\newacronym{los}{LOS}{line-of-sight}
\newacronym{gps}{GPS}{global positioning system}
\newacronym{siso}{SISO}{single-input single-output}
\newacronym[plural = EADFs, firstplural=effective aperture distribution functions (EADF)]{eadf}{EADF}{effective aperture distribution function}
\newacronym{aps}{APS}{angular power spectrum}
\newacronym[plural = UCAs, firstplural=uniform circular arrays (UCA)]{uca}{UCA}{uniform circular array}

\usepackage{footnote}
\makesavenoteenv{table}

\newcommand{\RNum}[1]{\uppercase\expandafter{\romannumeral #1\relax}}

\usepackage[table]{xcolor}
\usepackage{booktabs}
\usepackage{gensymb}

\usepackage[font=footnotesize]{caption}
\usepackage[font=footnotesize]{subcaption}

\begin{document}

\title{Real-Time Millimeter-Wave MIMO Channel Sounder for Dynamic Directional Measurements }


%

%
\author{
    \IEEEauthorblockN{C. Umit Bas, {\it Student Member, IEEE},
    Rui Wang, {\it Student Member, IEEE},
    Seun Sangodoyin, {\it Student Member, IEEE}, 
    Dimitris Psychoudakis, {\it Senior Member, IEEE},
    Thomas Henige,
    Robert Monroe, 
    Jeongho Park, {\it Member, IEEE},
    Jianzhong Zhang, {\it Fellow, IEEE}, 
    Andreas F. Molisch, {\it Fellow, IEEE}  
    } 
   
%
%

\thanks{C. U. Bas, R. Wang, S. Sangodoyin and  A.F. Molisch are with the Department of Electrical Engineering, University of Southern California (USC), Los Angeles, CA 90089-2560 USA.
D. Psychoudakis, T. Henige, R. Monroe, J. Zhang are with Samsung Research America, Richardson, TX, USA. J. Park is with Samsung Electronics, Suwon, Korea.}
\thanks{An earlier version of this work and initial results are presented in IEEE Vehicular Technology Conference Fall-2017 \cite{bas_2017_realtime} and in IEEE Vehicular Technology Conference Spring-2018 \cite{bas_2017_dynamic}, respectively.}
\thanks{This work is partially supported by the National Science Foundation and the National Institute of Standards and Technology.}
}


\maketitle
\IEEEpeerreviewmaketitle

\begin{abstract}
In this paper, we present a novel real-time multiple-input-multiple-output (MIMO) channel sounder for the 28~GHz band. Until now, most investigations of the directional characteristics of millimeter-wave channels have used mechanically rotating horn antennas. In contrast, the sounder presented here is capable of performing horizontal and vertical beam steering with the help of phased arrays. Due to its fast beam-switching capability, the proposed sounder can perform measurements that are directionally resolved both at the transmitter (TX) and receiver (RX) in 1.44 milliseconds compared to the minutes or even hours required for rotating horn antenna sounders. This not only enables measurement of more TX-RX locations for a better statistical validity but also allows to perform directional analysis in dynamic environments. The short measurement time combined with the high phase stability limits the phase drift between TX and RX, enabling phase-coherent sounding of all beam pairs even when TX and RX have no cabled connection for synchronization without any delay ambiguity. Furthermore, the phase stability over time enables complex RX waveform averaging to improve the signal to noise ratio during high path loss measurements. The paper discusses both the system design as well as the measurements performed for verification of the sounder performance. Furthermore, we present sample results from double directional measurements in dynamic environments.

\end{abstract}

\section{Introduction} \label{sec_intro}
The ever-growing need for higher data rates in wireless communications is motivating the use of previously unused spectrum.  Consequently, the millimeter wave (mm-wave) band has become a key area of interest for next generation wireless communication systems due to  the ample amount of bandwidth available at the frequencies higher than 6~GHz. It is now clear that mm-wave systems will be an essential component of 5th generation (5G) cellular networks \cite{rangan2014millimeter}.

The knowledge of true statistical characteristics of the propagation channel is imperative for designing and testing wireless systems \cite{Molisch_2010_book,Molisch_2016_eucap}. An accurate channel model is even more important for mm-wave bands, due to the unique propagation characteristics at those frequencies. It is anticipated that most of the future mm-wave systems will utilize beam-forming antenna arrays to overcome the higher path loss that occurs at higher frequencies. Hence the knowledge of angular spectrum and its temporal evolution are vital for the efficient design of such systems  \cite{roh2014millimeter}.

The basic operating principle of a channel sounder is to transmit a known waveform, so that suitable signal processing can deconvolve the transmitted signal from the received signal to acquire the impulse response of the channel. The sounding waveforms can be pulses \cite{Deparis_et_al_2005,Demir_2016_Arima}, pn-sequences \cite{Cassioli_2012,Ranvier_et_al_2007,Piersanti_et_al_2012,Zwick_2005_Wideband}, chirp signals  \cite{Hakegard_et_al_2016_PIMRC,Chandra_et_al_2016_NTMS}, or multitone sequences \cite{Conrat_et_al_2006,Peter_and_Keusgen_2007} (see also Sec. II.A). Most of the existing directional channel sounders for indoor mm-wave systems are based on vector network analyzers (VNAs), which use slow chirp or frequency stepping, combined with virtual arrays (mechanical movement of a single antenna along a track). Such setups cannot operate in real-time, and need cabled connections between transmitter (TX) and receiver (RX); they are thus mostly used for static indoor channel measurements \cite{Smulders_and_Wagemans_1992_PIMRC,Gustafson_et_al_2011_SAGE, Gustafson_and_Tufvesson_2012,Fu_et_al_2013,Blumenstein_2017_invehicle}. 
 For outdoor measurements, the prevalent method for directionally resolved measurements is based on mechanically rotating horn antennas \cite{Rappaport_et_al_2013_AP,Rappaport_et_al_2015_TCom,hur_synchronous_2014,Kim_et_al_2015,ko2017millimeter,MacCartney_2017_flexible,du2018suburban}, whose operating principle is sketched in Fig. 1: a (single-input-single-output) channel sounder is connected to horn antennas that are mechanically rotated. For each pair of directions at TX and RX antennas, the sounder measures the impulse response. Note, however, that mechanical rotation requires measurement durations on the order of {\em 0.5-5~hours} for one measurement. One exception is the channel sounder presented in \cite{du2018suburban} which can perform $360^\circ$ sweeps in 200~ms but this is only for the sweep on one side (TX), and only for narrowband measurements. Consequently, the measurements reported at mm-wave bands either investigated angular characteristics but ignored the channel's temporal variation, or focused on dynamic channel characteristic \cite{He_2018_influence,MacCartney_2017_rapid} without considering angular characteristics.

\begin{table*}[t]\centering
	\caption{mm-wave channel sounder comparison}
	\scriptsize
	\renewcommand{\arraystretch}{1}
	\begin{tabular}{l|p{2.3cm}|p{2.3cm}|p{2.3cm}|p{2.3cm}|p{2.3cm}}
		\hline
		\textbf{Specifications} & \textbf{USC \cite{bas_2017_realtime}} & \textbf{NYU \cite{MacCartney_2017_flexible}} & \textbf{NIST \cite{Papazian_et_al_2016,sun2017design}} & \textbf{Durham\cite{Salous_2016_EuCAP,Salous_2016_twc} }& \textbf{Keysight \cite{keysight_sounder,Wen_2016_mmwave}} \\ \hline \hline
		Center Frequency & 27.85~GHz &28, 38, 60, 73~GHz & 28.5, 60~GHz & 30, 60, 90~GHz & Up to 44~GHz \\
		Bandwidth &  400~MHz & 1~GHz & 2~GHz &  3, 6, 9~GHz & 2~GHz\\
		TX EIRP & 57.1~dBm & 54.6~dBm & 51.5~dBm & 36.7~dBm & 23~dBm + Ant. Gain \\
		Array Type & Switched Beam & Rotating Horn & Switched Horn  & Switched & TX: Switched RX: Parallel\\
		Array Size & 16$\times$16 & NA &  16$\times$8 & 8$\times$8 &8$\times$8 \\
		RX Antenna/Beam Gain & 19.5~dBi & 24.5~dBi & 18.5~dBi & 20.7~dBi & -  \\
		Direction switching speed & 2~$\mu s$ & $>$seconds & 4~$\mu s$ & -  & - \\
		TX-RX Synchronization & GPS-Rubidium Ref. & Rubidium Ref. & GPS-Rubidium  Ref. & Rubidium  Ref. & GPS-Rubidium  Ref.\\
		\hline
	\end{tabular} \label{tab:comparison}
\end{table*}

\begin{figure}
\centering
\begin{subfigure}{.48\linewidth}
  \centering
  \includegraphics[width=1\linewidth]{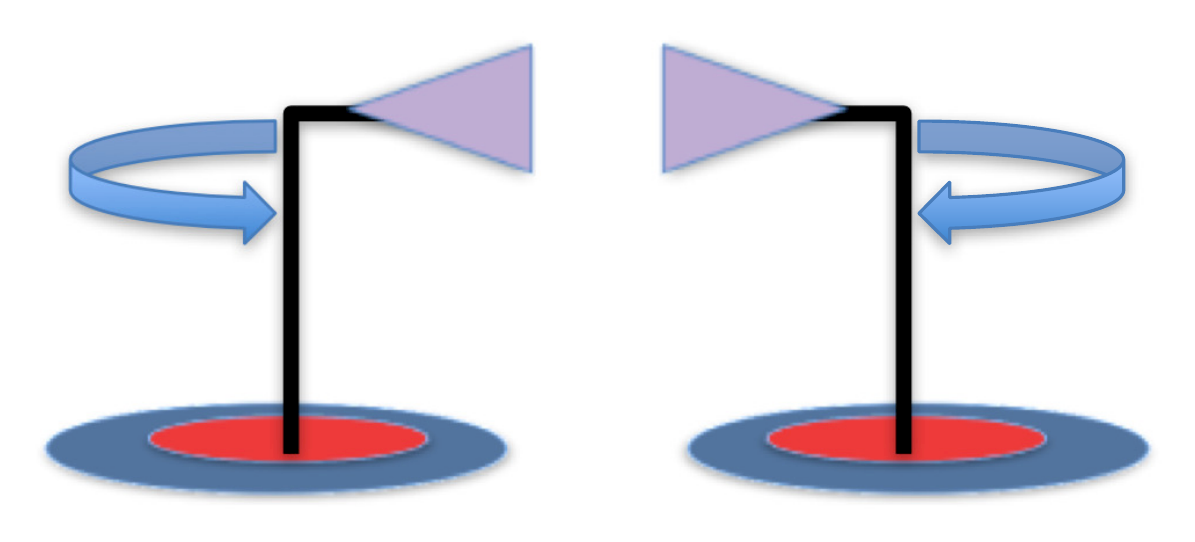}
    \caption{Rotating horn antenna}\label{fig:horns}
\end{subfigure}%
\begin{subfigure}{.48\linewidth}
  \centering
  \includegraphics[width=1\linewidth]{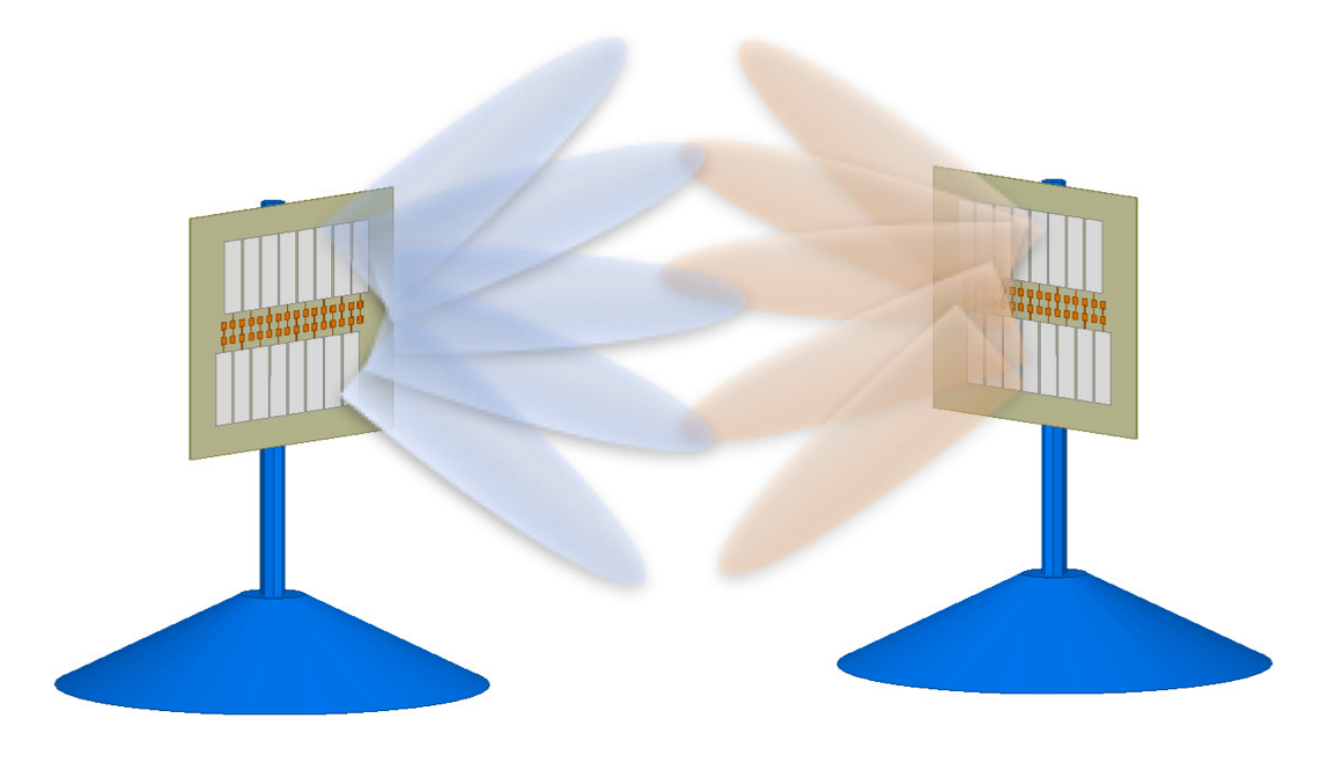}
    \caption{Phased array}\label{fig:phased_array}
\end{subfigure}
\caption{Approaches for directional mm-wave channel sounding}
\label{fig:test}
\end{figure}

In this paper, we present a novel real-time channel sounder setup that can perform directionally resolved measurements within 90$^\circ$ sectors at the TX and the RX. The setup presented in this paper operates at 28~GHz, although its design principles can be applied to other mm-wave bands as well. It is based on an approach of electronically-switched beams. By using arrays of antennas with phase shifters, we form beams into different directions at both the TX and the RX as shown in Fig. \ref{fig:phased_array} instead of using the rotating horn approach in Fig. \ref{fig:horns}. With a control interface implemented in field programmable gate array (FPGA), we are capable of switching from one beam to another in less than 2~$\mu s$. Thus, the same effect as in rotating horn antennas (pointing beams in different directions) can be achieved in much shorter time.

Since our proposed setup decreases the measurement time for a single multiple-input-multiple-output (MIMO) snapshot from hours to milliseconds, it allows the collection of tens of thousands of measurement points in a single measurement campaign, i.e., many orders of magnitude more than with rotating horns. Furthermore, since all TX and RX antenna pairs can be measured within the coherence time of the channel, the developed sounder is suitable for directionally-resolved measurements in dynamic environments. Hence, the data acquired with this channel sounder can provide valuable insights into the effects of moving vehicles or pedestrians on the angular channel statistics which are crucial inputs for systems utilizing beam-forming - such as mm-wave 5G systems. 
Due to the capability of streaming the received data to a large storage, the temporal evolution of multipath components (MPC) can be tracked over long times.

The short measurement time together with careful RF design to reduce phase noise limits the relative phase drift between the local oscillators (LO) of the TX and RX within a single MIMO snapshot even without a cabled connection for synchronizing the clocks. The phase coherence not only eliminates any delay uncertainty between measurements for different TX-RX directions but also enables complex RX waveform averaging to improve the signal to noise ratio for high path-loss measurement scenarios. Furthermore, it can provide one of the necessary conditions for employing high resolution parameter extraction (HRPE), such as RIMAX \cite{rui_Enabling_2018} at mm-wave frequencies.

Table \ref{tab:comparison} lists the state-of-the-art mm-wave channel sounders. Parallel to our work, three other groups have developed real-time capable mm-wave sounders: Ref. \cite{Papazian_et_al_2016,sun2017design} recently presented a sounder based on an array of horn antennas combined with an electronic switch at the RX and a single TX antenna. This design is capable of faster sounding, and is similar in spirit to our approach. There are two important differences to our setup: (i) it has significantly less equivalent isotropically radiated power (EIRP) even with the horn antennas (ii) the use of mechanically arranged horns limits the flexibility compared to our sounder, which can reconfigure the beam-patterns through software. The sounder described in Ref. \cite{Salous_2016_EuCAP,Salous_2016_twc} is a multi-band and multi-antenna channel sounder operating at carrier frequencies of 30~GHz, 60~GHz and 90~GHz with 8$\times$8, 2$\times$2 and 2$\times$2 antenna arrays, respectively. The MIMO operation is realized by employing switches at the intermediate frequency (IF) along with parallel frequency conversions. However, nominal TX output powers are limited to 16~dBm at 30~GHz, 7~dBm at 60~GHz and 4~dBm at 90~GHz. Furthermore, the MIMO order is lower than the 16$\times$16 achieved in our setup.  Finally, Ref. \cite{Wen_2016_mmwave} presents a sounder with 4 TX antennas multiplexed with a switch and 4 RX antennas with 4 down-conversion chains. Similar to \cite{Salous_2016_EuCAP}, the TX power is limited to 24~dBm. 

Our channel sounder has been used to measure different aspects of the wireless propagation channel at \SI{28}{GHz}. In \cite{bas_2017_microcell}, we presented path loss and root mean square delay spread for a suburban micro-cell environment. Ref. \cite{bas_2017_O2I} investigated the effects of outdoor to indoor penetration on path loss, penetration loss, delay spread and angular spread statistics for two different types of housing. In \cite{wang_2017_stationarity}, we presented the first measurement results for the stationarity region of MIMO mm-wave channels, which were based on over 20 million channel impulse responses measured on continuous routes. Ref. \cite{choi_human_2018} investigates the human body shadowing and its effects on the angular spectrum. Finally, \cite{bas_2017_dynamic} discussed the first measurement campaign for a {\em dynamic double-directionally resolved} measurement campaign at mm-wave frequencies for an outdoor micro-cellular scenario. The real-time measurement capability of the setup enabled these measurements, which are not possible with rotating horn antenna channel sounders.

The rest of the paper is organized as follows. Section \ref{sec_design} discusses the proposed channel sounder setup.  Section \ref{sec_verification} explains the measurements that verify system performance. Section \ref{sec_post} discusses the post processing of the measurement data. Section \ref{sec_meas} presents results from a dynamic directional measurement campaign. Finally Section \ref{sec_conc} summarizes results and suggests directions for future work.

\section{Channel Sounder Design} \label{sec_design}

This section describes the channel sounder design. As for any channel sounder, a known waveform is generated in baseband, up-converted to passband and transmitted over the air. The received waveform is down-converted, sampled, and stored, for further post processing. The channel impulse response is extracted from the shape of the received signal. In order to extract the directional channel properties, repetitions of the sounding signal are sent into (and received from) different directions, where the transmit and receive beams are generated by phased arrays. The following sections describe these aspects in more detail. 

\begin{figure}\centering
	\centering\includegraphics[width=0.6\linewidth]{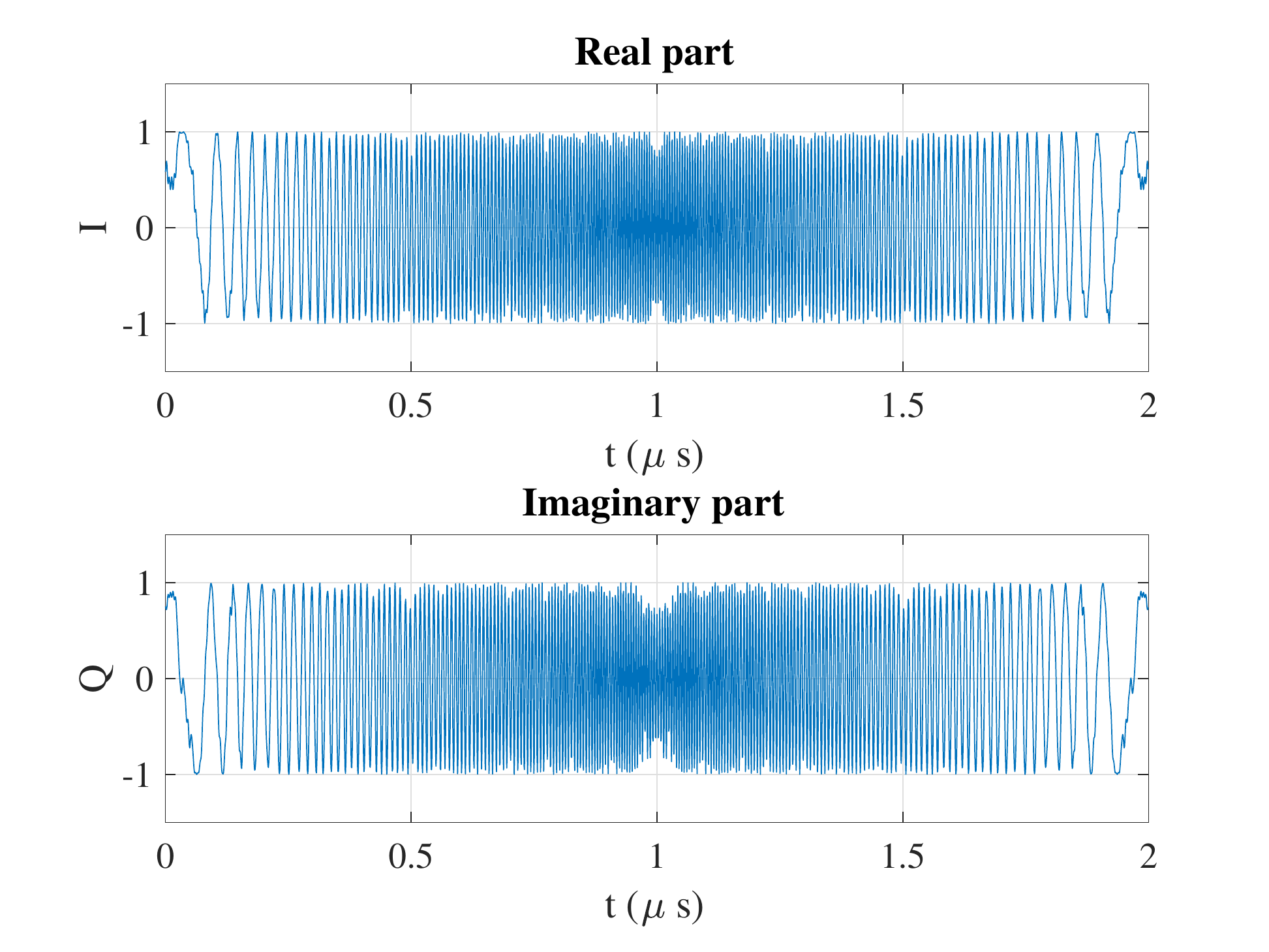}\caption{In-phase and Quadrature components of the baseband signal, with a sampling rate of 1.25~GSps}\label{fig:sounding_IQ}
    \vspace{0.1in}
	\centering\includegraphics[width=0.6\linewidth]{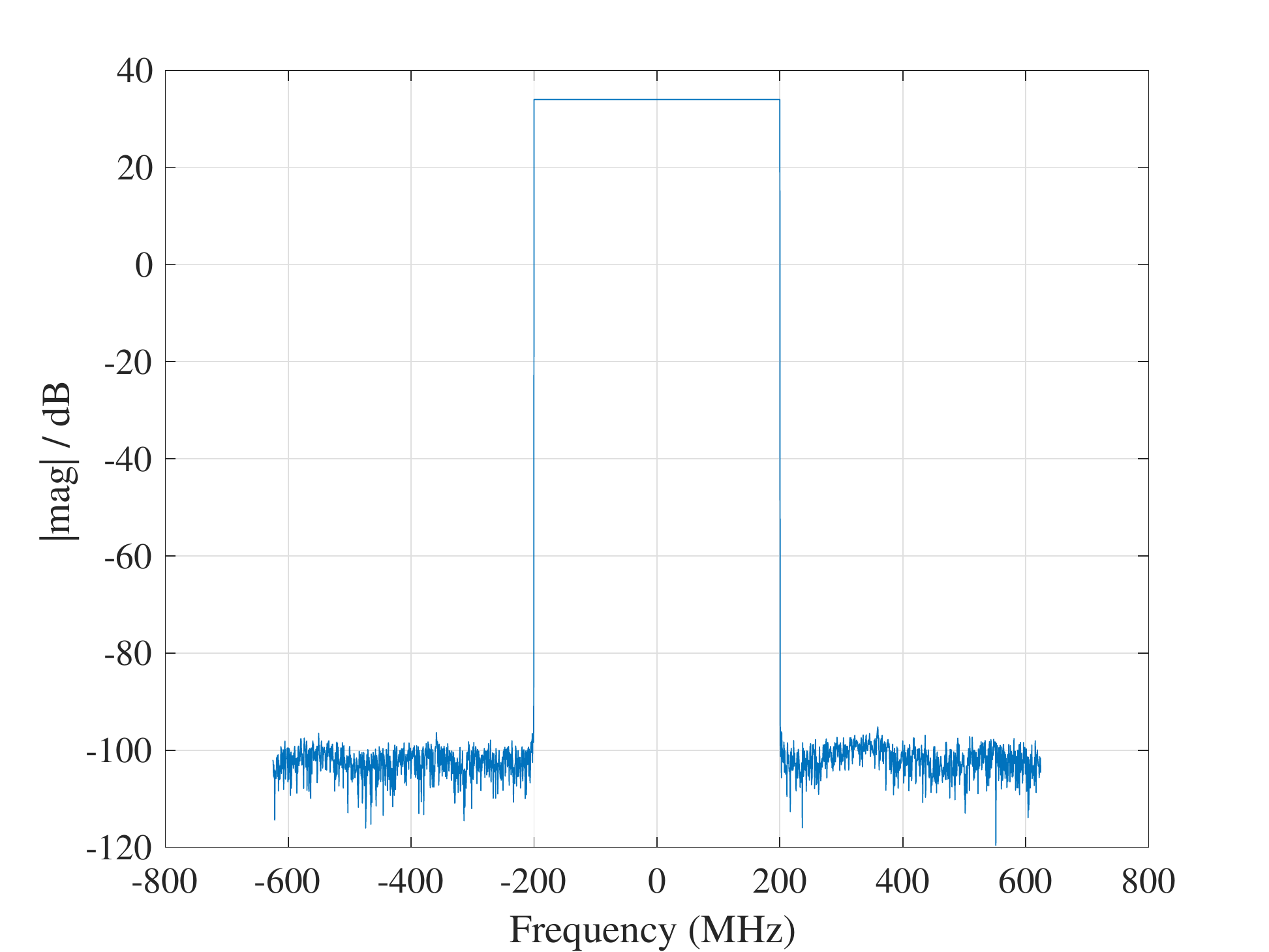}\caption{Spectrum of the baseband multi-tone sounding signal with 801 tones spaced by 500~kHz}\label{fig:sounding_bb_hf}
    \vspace{0.1in}
	\centering\includegraphics[width=0.6\linewidth]{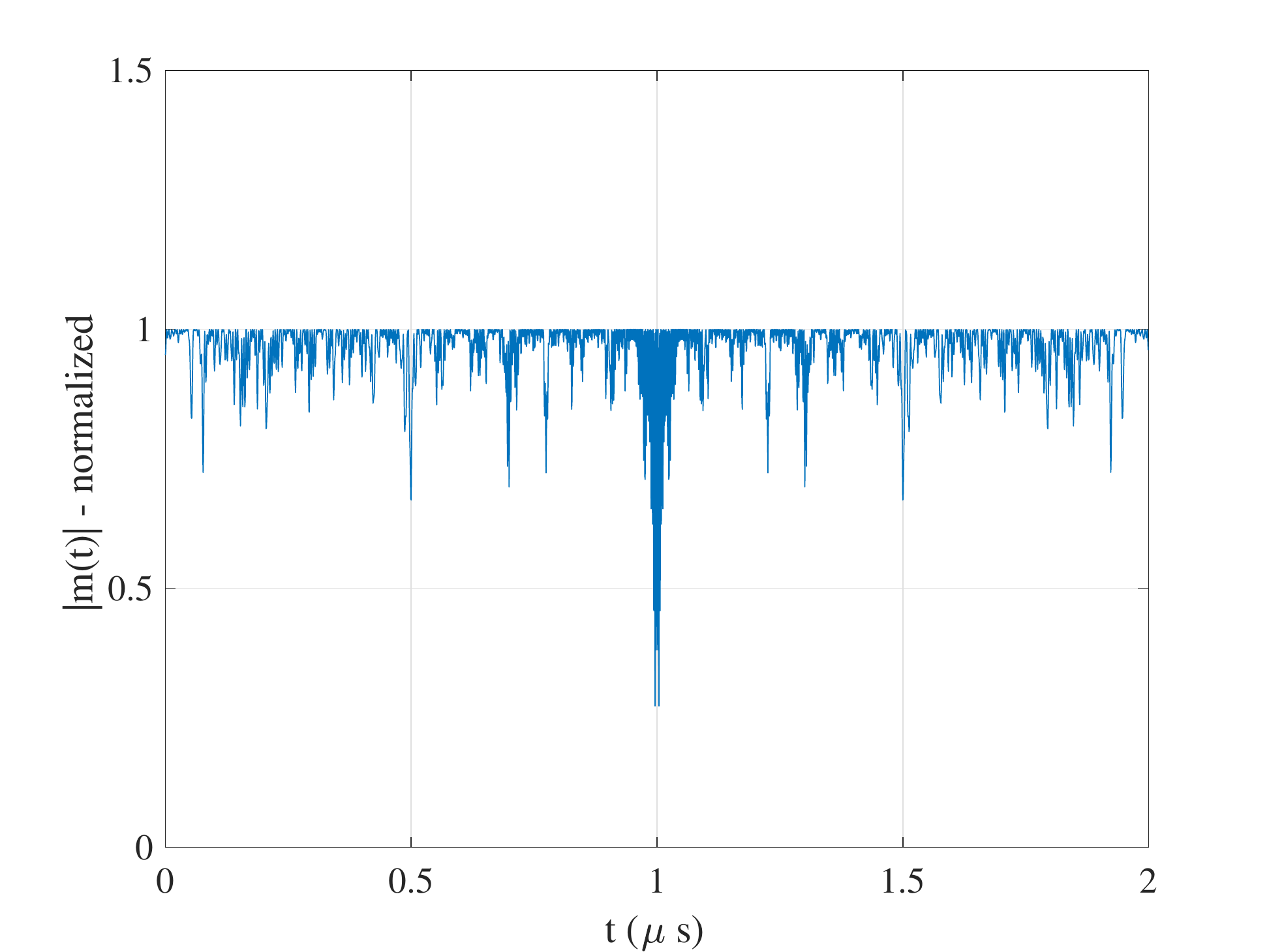}\caption{Normalized magnitude of the sounding signal}\label{fig:sounding_abs}
\end{figure}

\subsection{Sounding Waveform}\label{sec:wave}

The baseband sounding waveform used throughout this work is a multi-tone waveform. Multi-tone waveforms are a good fit for channel sounding measurements since they can be flat in both in the time and the frequency domain. For mm-wave sounders, they are used e.g., in \cite{Conrat_et_al_2006} and in \cite{Peter_and_Keusgen_2007}. The sounding waveform can be represented as:
\begin{equation}
  m(t)= \sum_{n=-N}^{N} e^{j (n 2\pi \Delta f t + \theta_n)}
\end{equation}
where $\Delta f$ is the tone spacing, $2N+1$ is the number of tones and $\theta_n$ is the phase of the tone $n$. Fig. \ref{fig:sounding_IQ}  and \ref{fig:sounding_bb_hf} show the in-phase and quadrature components, and the spectrum of the complex baseband waveform $ m(t)$, respectively. The spectrum of the sounding waveform has a flat-top providing the same signal-to-noise ratio at all frequency tones. As suggested in \cite{Friese1997multitone}, the values of $\theta_n$ can be modified to achieve a low peak to average power ratio (PAPR). Fig. \ref{fig:sounding_abs} shows the normalized amplitude of the waveform, which has a PAPR of $0.4$ dB, allowing us to transmit with power as close as possible to the 1 dB compression point of the power amplifier without driving it into saturation. Note that while Zadoff-Chu sequences, which are, e.g., used in LTE, provide PAPR=1 under idealized circumstances, this does not hold true for {\em filtered, oversampled} sequences, which are relevant here.  Our sequences outperform Zadoff-Chu by more than 1 dB.

\begin{figure}\centering
       \centering\includegraphics[width=1\linewidth, viewport=15 15 640 315, clip=true] {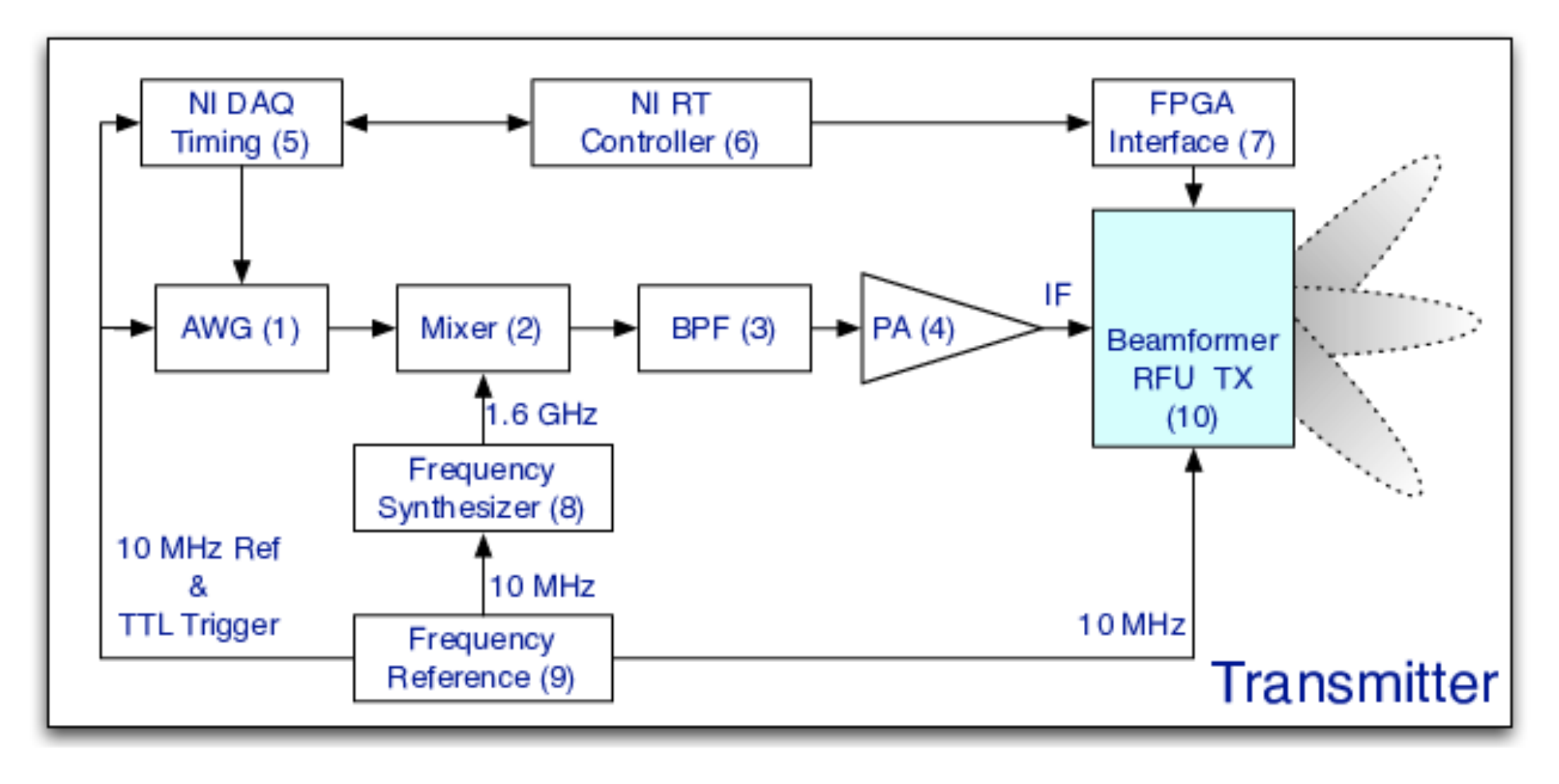}\caption{Channel sounder TX block diagram}\label{fig:TX} 
    \vspace{0.2in}
	    \centering\includegraphics[width=1\linewidth, viewport=15 15 640 315, clip=true] {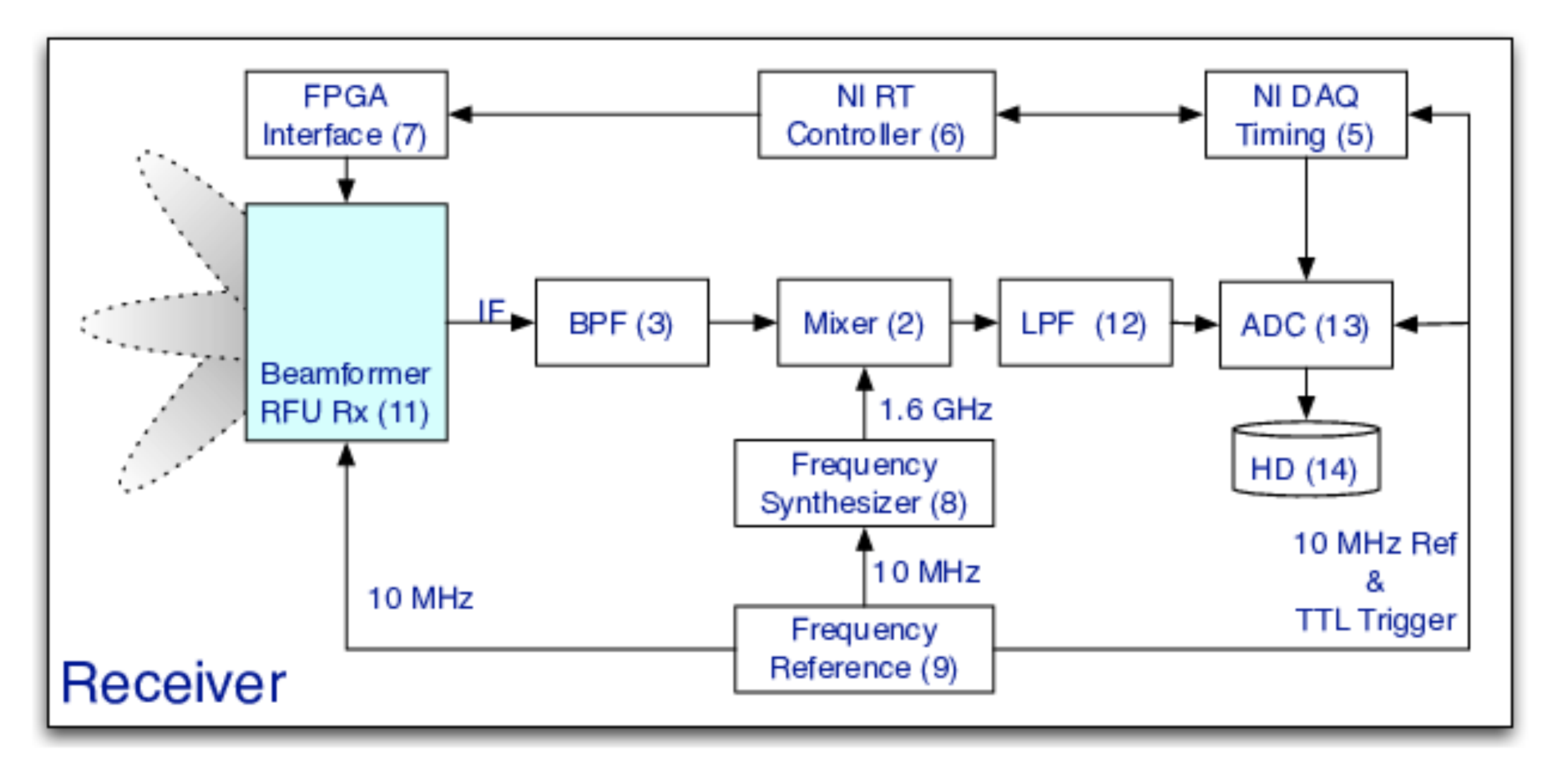}\caption{Channel sounder RX block diagram}\label{fig:RX}
\end{figure}

\subsection{Sounder Hardware}

The developed channel sounder is a beam-switched multi-carrier setup with 400 MHz instantaneous bandwidth. Figs. \ref{fig:TX} and \ref{fig:RX} show the block diagrams for the TX and RX, respectively. The descriptions of the components shown in the block diagrams are listed in Table \ref{tab:parts}. 

\begin{table*}\centering \caption{Descriptions of the components in the TX and RX block diagrams}
	\renewcommand{\arraystretch}{1}
	\scriptsize
	\begin{tabular}{|c|l|l|} 
		\hline
		\textbf{Unit Number} & \textbf{Unit Name} & \textbf{Unit Description} \\ \hline
		(1) & Agilent N8241A & AWG: 15-bit, 1.25-GSps  \\ 
		(2) & Mini-Circuits ZEM-M2TMH+ & Double-balanced mixer, conversion loss $\le$ 7~dB  \\ 
		(3) & KL Microwave 11ED50-1900/T500-O/O & Band pass filter, insertion loss $\le$ 0.4~dBa, pass band: 1.65-2.15~GHz  \\ 
		(4) & Wenteq ABP1500-03-3730 & Amplifier: 0.5-15~GHz, 37~dB gain  \\ 
		(5) & National Instruments PXIe-6361 & Multifunction I/O Module: 32-bit counters at 100~MHz clock \\
		(6) & National Instruments PXIe-8135 & 2.3 GHz Quad-Core PXI Controller\\
		(7) & National Instruments PXIe-7961R & Virtex-5 SX50T FPGA PXI FPGA Module with LVDS and LVTTL I/O\\
		(8) & Phase Matrix FSW-0020 & Frequncy Synthesizer: 0.2-20~GHz\\
		(9) & Precision Test Systems GPS10eR & GPS-disciplined Rubidium Reference: Allan deviation $\le$ 1.5e-12 \\
		(10) & Samsung 28 GHz RFU TX & Up-converter and phased array \\
		(11) & Samsung 28 GHz RFU RX & Down-converter and phased array\\
		(12) & Low Pass filter & Cut off frequency 500~MHz  \\ 
		(13) & National Instruments PXIe-5160 & Digitizer: 10-bit, 1.25-GSps   \\
		(14) & National Instruments HDD-8265 & Raid Array: 6~TB capacity, 700~MBps read/write speed  \\
		\hline
	\end{tabular}\label{tab:parts}
\end{table*}

\begin{figure}\centering
    	\centering\includegraphics[width=0.6\linewidth, viewport=38 180 550 600, clip=true]{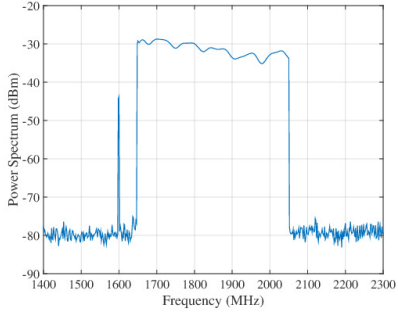}\caption{Spectrum of the sounding signal at IF which is the input of the TX RFU. The sounding signal consists of 801 tones with a total bandwidth of 400~MHz centered at 1.85~GHz. The LO leakage at 1.6~GHz is less than -40~dBm} \label{fig:sounding_sig}
\end{figure}

Fig. \ref{fig:TX} shows the block diagram for the TX. A 15-bit, 1.25-GSps arbitrary waveform generator (AWG) generates the baseband sounding signal that has equally spaced 801 tones covering the frequency range from 50 MHz to 450 MHz. After the baseband signal is generated, a mixer up-converts it with a local oscillator (LO) frequency of 1.6 GHz. Since we only utilize the upper sideband of the up-converted signal as the intermediate frequency (IF) input of the beam-former radio frequency unit (BF-RFU), a band-pass filter suppresses the LO leakage of the mixer and the lower sideband. After the band-pass filter and the pre-amplifier, we obtain an IF signal with 400 MHz bandwidth centered at 1.85 GHz as seen in Fig. \ref{fig:sounding_sig}. The peak seen at the frequency of 1.6 GHz is due to residual LO leakage. It is at least 10 dB lower than the sounding tones and approximately 40 dB lower than the total power of the multi-tone signal, hence it does not affect the measurements in a significant manner.

\begin{figure}\centering
	\centering\includegraphics[width=1\linewidth, viewport=15 15 585 380, clip=true]{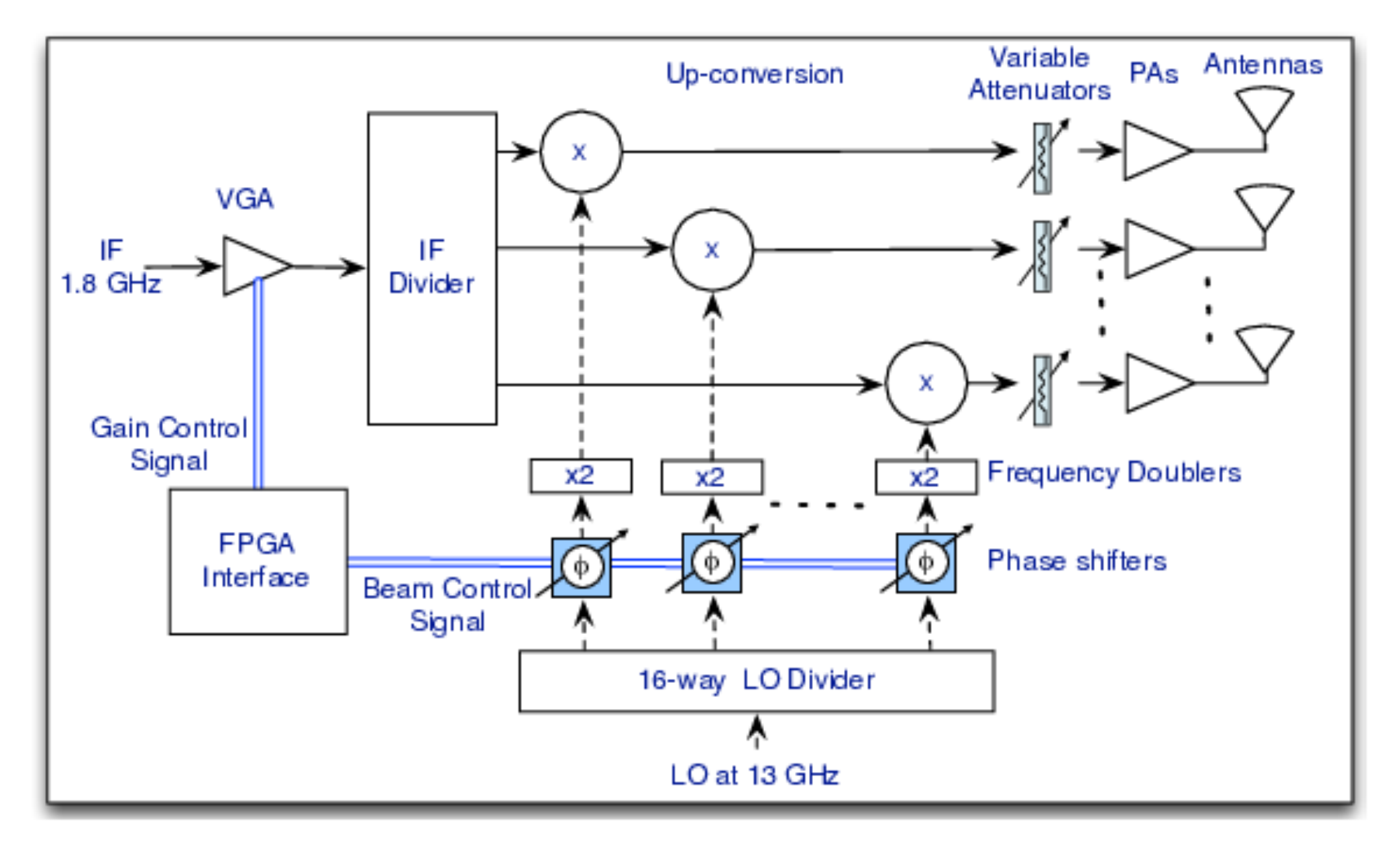}\caption{Block diagram for TX RFU which performs gain control, up-conversion and beam-forming }\label{fig:RFU}
    \vspace{0.1in}
	\centering\includegraphics[width=0.3\linewidth]{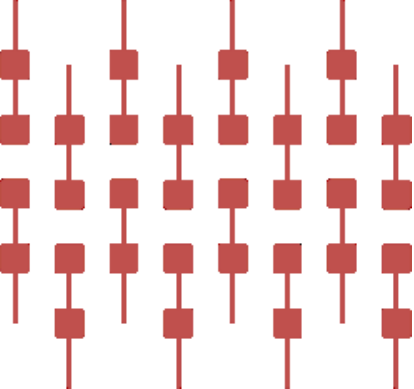}\caption{Phased array with 8 by 2 antenna elements, each element consists of two patches, same array structure is used both in the RX and TX}\label{fig:antenna}
\end{figure}

Finally, the IF signal is fed to the BF-RFU, whose simplified block diagram is shown in Fig. \ref{fig:RFU}. Similar to the approach in \cite{Psychoudakis_2016_mobile}, within the BF-RFU, this IF signal is split into 16 identical signals, which are fed into 16 RF chains. Each chain performs an up-conversion to 27.85 GHz with dedicated amplifiers and phase shifters to perform beam-forming. The antenna array is made of 8 by 2 antenna elements, each element consists of two patches as shown in Fig. \ref{fig:antenna}. The subarray spacing is \SI{5.6}{mm} ($0.52\lambda$ at 28 GHz) horizontal and \SI{12.5}{mm} ($1.16\lambda$ at 28 GHz) in vertical. The BF-RFU allows to perform 90$^\circ$ horizontal beam steering and 60$^\circ$ vertical beam steering with 5$^\circ$ steps. The phase shifters can be configured via a control interface, thus allowing an adaptation of the beam-shapes between measurement campaigns, see also Section \ref{sec:operation}. The phase shifters used in the phased arrays have a resolution of $5.625^\circ$, the frequency doubler also doubles that to $11.25^\circ$. Due to the design, the number of beams between which we can switch can be larger than the number of antennas (subarrays). Another key feature of the RFUs is their low phase noise, see Section \ref{sec:phase}.

Another key aspect of the sounder is the high EIRP. By design, the power amplifiers of all RF chains are powered up continuously (except for breaks during the during the switching between beams). This is a significant difference to sounders where different directions are excited by a circular array of horn antennas, since in that case one power amplifier (PA) (corresponding to one beam) is active at one time. For the same specifications of the PA, the transmit power in our design is thus higher by 12~dB (factor 16). Since it is desirable to stay in the power amplifiers' linear regime of operation while maximizing the transmitted power, the output power is set 3.4~dB (PAPR plus 3~dB) less than the 1dB compression point of the amplifiers. The 1 dB compression point of the individual amplifier is $31$~dBm per amplifier, however, thanks to $16$ power amplifiers used in parallel, the total output power is $39.6$~dBm with $3.4$~dB back off. After $2$~dB feed loss and taking into account the $19.5$~dBi antenna gain (\SI{12}{dBi} array gain plus \SI{7.5}{dBi} gain of 2 element subarrays), we achieve $57.1$~dBm EIRP at the output of the TX array. 

Similar to the TX RFU, the RX RFU also consists of a 16-element beam-forming antenna array. After low-noise amplification and combining of the received power, there is an additional step of automatic gain control to utilize the dynamic range of the RX to its limits. The RX gain control has a range of \SI{60}{dB} with \SI{0.5}{dB} steps. For each MIMO snapshot, the RX controller estimates the power received from the TX-RX beam pair with the highest power, then adjusts the RX gain accordingly to ensure that the received waveforms always have the same power level prior to the IF to baseband conversion. The gain control is performed prior to each MIMO snapshot, and the gain levels are recorded along with the measurement data, GPS coordinates, and the coordinated universal time (UTC) from the GPS clock. At the output of the RX BF-RFU, the received IF signal is filtered, down-converted back to baseband and finally sampled by a 1.25-GSps 10-bit digitizer. The digitizer streams the sampled data to a redundant array of independent disks (RAID) with a rate up to 700~MBps and stores the data for post processing as shown in Fig. \ref{fig:RX}. 

The receiver sensitivity at room temperature with 400~MHz bandwidth is -83~dBm. Combined with the RX beam-forming and antenna gains (19~dB), and TX EIRP (57~dBm), we achieve a measurable path loss of $159$~dB without any averaging or spreading gain, see Section \ref{sec:dynamic range} for further discussion.

\subsection{Sounder Operation}\label{sec:operation}

Both the TX and the RX are controlled with LabVIEW scripts running on National Instruments PXIe controllers. The beam steering and gain of the variable gain amplifiers at the BF-RFUs are controlled via an FPGA interface with a custom designed control signaling protocol implemented in LabVIEW FPGA. This interface allows us to switch between any beam setting or gain setting in less than 2~$\mu s$. Consequently, the proposed sounder can complete a full-sweep a million times faster than a virtual array. All beam pairs can be measured without retriggering the digitizer or the AWG. This avoids triggering jitter which would create uncertainty in the absolute delay of the paths observed in different beams. TX and RX have no physical connections and they are synchronized with GPS-disciplined Rubidium frequency references. These references provide two signals for the timing of the setup; a 10~MHz clock to be used as a timebase for all units and 1 pulse per second (PPS) signals aligned to UTC.

Given the measurement period, hardware counters in the NI DAQ Timing modules count the rising edges of the 10~MHz and trigger the rest of the units at given times. These counters are in turn triggered by the 1~PPS. Since the 1~PPS signals in the TX and RX are both aligned to the UTC, they operate synchronously without requiring any physical connections. More importantly, the AWG, the ADC, the frequency synthesizers and the BF-RFUs are disciplined with the 10~MHz signal provided by these frequency references, so that they maintain phase stability during the measurements, which is essential for accurate measurement results \cite{almers2005effect}. These references also provide GPS locations which are logged along with the measurement data.

\begin{table}[tbp]\centering\setlength\belowcaptionskip{-20pt}
	\caption{Channel sounder specifications}
	\renewcommand{\arraystretch}{1}
	\begin{tabular}{l|c}
		\hline
		\multicolumn{2}{c}{\textbf{Hardware Specifications}} \\ \hline \hline
		Center Frequency & 27.85 GHz\\
		Instantaneous Bandwidth & 400 MHz\\
		Antenna array size & 8 by 2 (for both TX and RX) \\
		Horizontal beam steering & -45$^\circ$ to 45$^\circ$  \\
		Horizontal 3dB beam width & 12$^\circ$\\
		Vertical beam steering & -30$^\circ$ to 30$^\circ$  \\
		Vertical 3dB beam width & 22$^\circ$\\
		Horizontal/Vertical steering steps & 5$^\circ$\\
		Beam switching speed & 2 $\mu s$ \\
		TX EIRP & 57 dBm \\
		RX noise figure & $\le$ 5 dB \\ 
		ADC/AWG resolution & 10/15-bit \\
		Data streaming speed & 700 MBps \\ \hline
		\multicolumn{2}{c}{\textbf{Sounding Waveform Specifications}} \\ \hline \hline
		Waveform duration & 2 $\mu s$ \\
		Repetition per beam pair & 10 (1 for dynamic) \\
		Number of tones & 801 \\
		Tone spacing & 500 kHz \\
		PAPR & 0.4 dB \\ 
		Total sweep time\footnotemark & 14.44 ms (400 $\mu s$ for dynamic)  \\ \hline  
	\end{tabular} \label{tab:specs}
\end{table}
\footnotetext{{For 1 elevation and all 19$\times$19 RX-TX azimuth beam combinations (10$\times$10 for dynamic)}}

Finally, Table \ref{tab:specs} summarizes the hardware and sounding waveform specifications. While this is the configuration used throughout this paper, the sounding waveform can be modified without any significant changes in the hardware. Equally importantly, the beams can be configured by modifications of the FPGAs in the RFUs. This enables, for example, to get fast scanning in azimuth only for situations where we know that MPCs are mainly incident in the horizontal plane, while a slower sweep through azimuth and elevation can be implemented for other cases. This is a significant advantage compared to setups with switched horn arrays, which cannot be reconfigured without extensive mechanical modifications, recabling, and/or switches with larger sizes than the actually used number of antennas. For example, while investigating relatively more stationary environments, we employ more waveform averaging and a higher number of beams \cite{bas_2017_microcell,bas_2017_O2I} in comparison to faster measurements performed in the more dynamic environments \cite{wang_2017_stationarity,bas_2017_dynamic}.

\section{System Verification} \label{sec_verification}

\subsection{Beam-Steering}

\begin{figure}[tbp]
	\centering\includegraphics[width=0.6\linewidth]{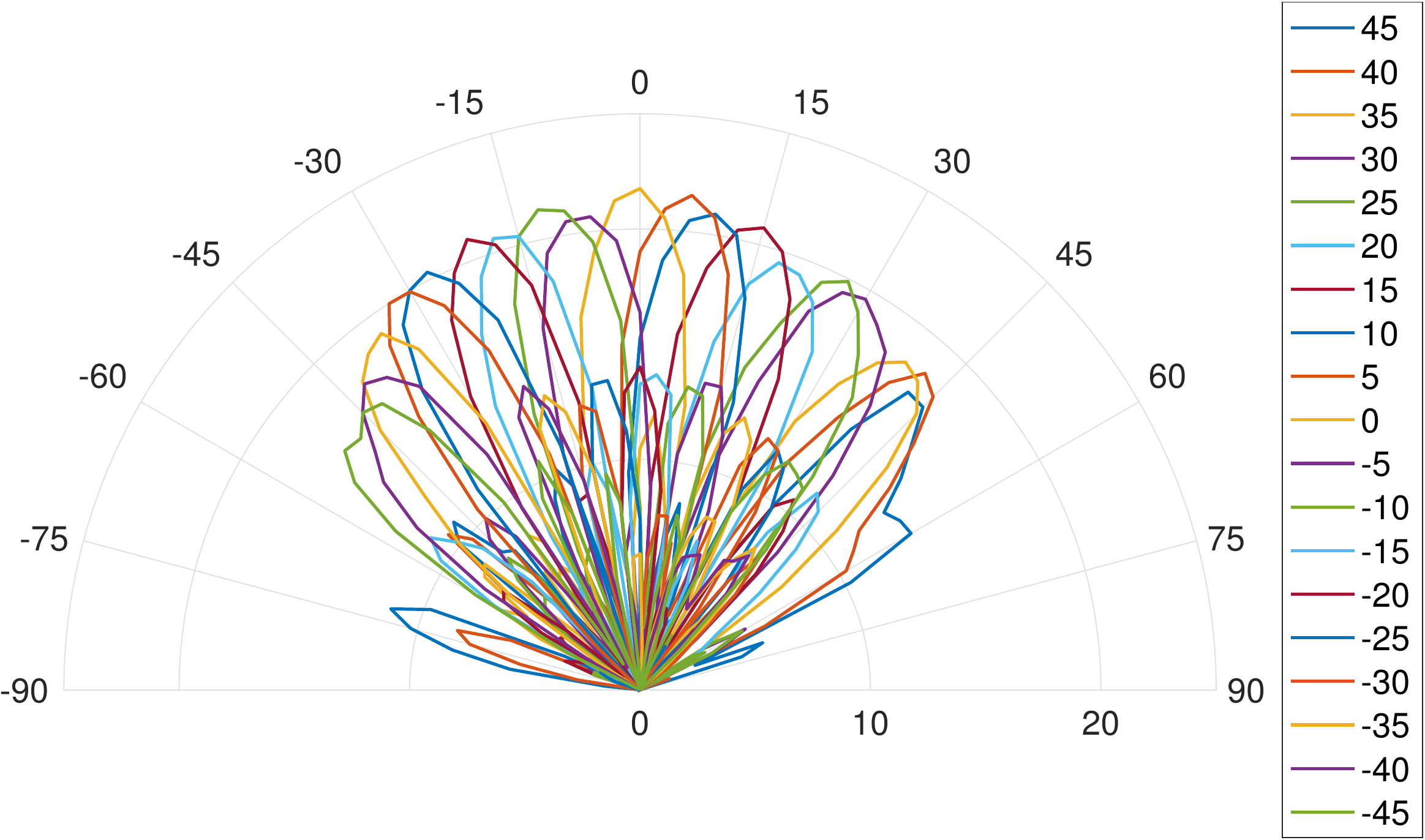}\caption{Measured azimuth patterns (dB) of the beams steering from -45$^\circ$ to 45$^\circ$, at the elevation angle of 0$^\circ$ }\label{fig:pattern}
\end{figure}

\begin{figure}[tbp]  
	\centering\includegraphics[width=1\linewidth, viewport=100 25 1140 345, clip=true]{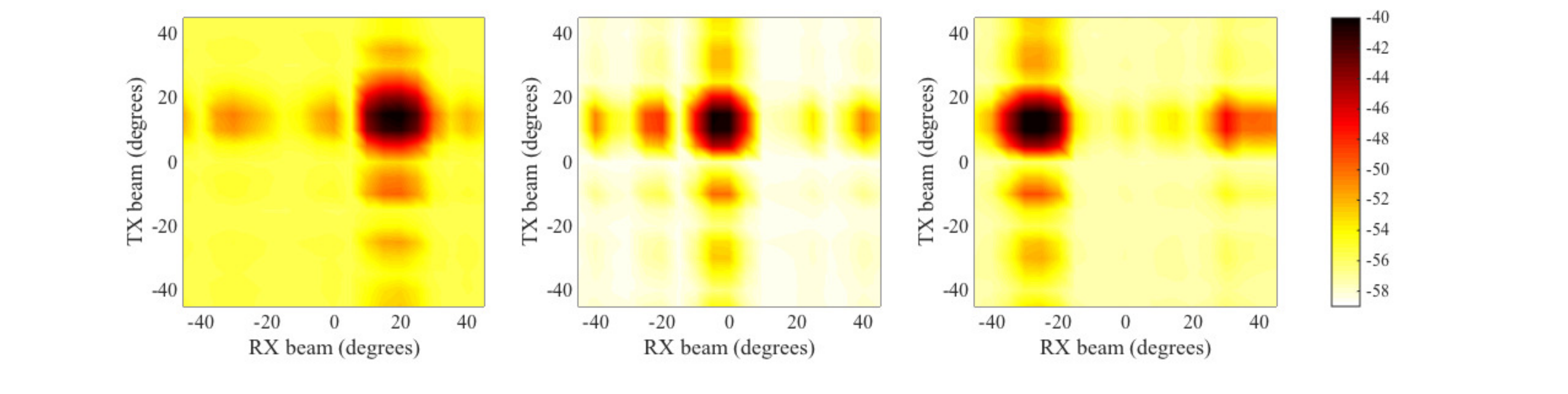}\caption{Received power  (dB)  vs TX and RX beam pairs when LOS component is at RX angle $x$ and TX angle $y$ and $(x,y)=\{(20,15),(-5,15),(-30,15)\}$ }\label{fig:beams}
\end{figure}

Firstly, we show the patterns of the beams formed by the TX BF-RFU, the results are representative for the RX side as well. Fig. \ref{fig:pattern} shows the beam patterns for all azimuth beams for the TX. In azimuth, 19 beams cover the range -45$^\circ$ to +45$^\circ$ with 5$^\circ$ steps. All beams have approximately 12$^\circ$ 3 dB beam-width and the side lobes are -10 dB or less relative to the main beam. Consequently, we investigate the directional channel characteristics by utilizing the beam directionality. Fig. \ref{fig:beams} shows the received power for all horizontal TX-RX beam pairs, where the LOS component is at x degree (relative to the 0 degree point of the physical array of the TX, and y degree relative to the RX array). Since the beams are overlapping, significant power is recorded at several beam positions. While this effect naturally limits the angular resolution when directions are determined based on the "strongest beam" only (similar to horn antennas), the effect is actually desired for HRPE algorithms which usually depend on the relative phase shift of a MPC received by different antennas or beams as in our case.

\begin{figure}[tbp]\centering
    \centering\includegraphics[width=0.6\linewidth, clip=true]{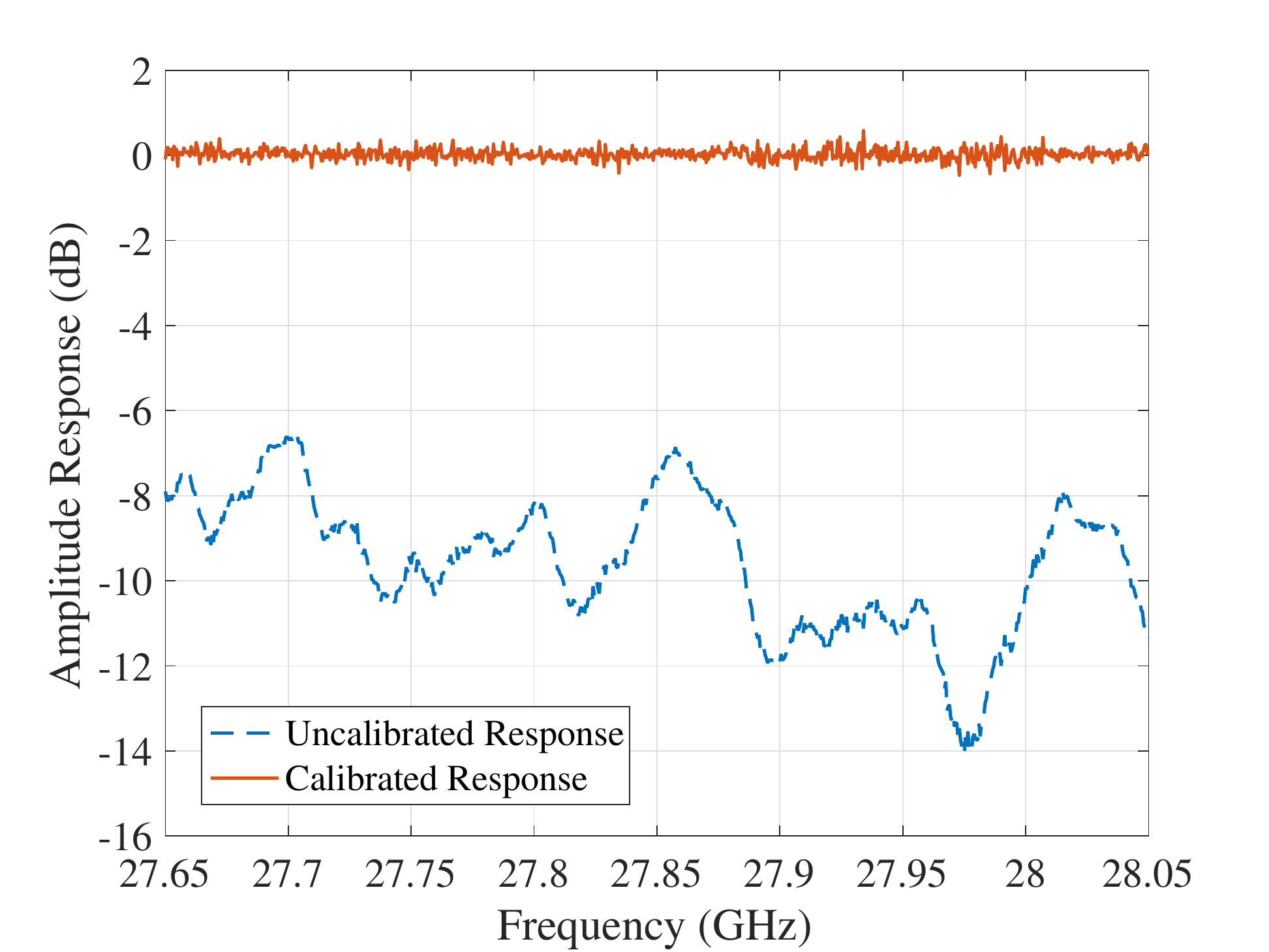}\caption{Calibrated and uncalibrated amplitude responses of the channel sounder}\label{fig:frequency}
    \vspace{0.1in}
	\centering\includegraphics[width=0.6\linewidth, clip=true]{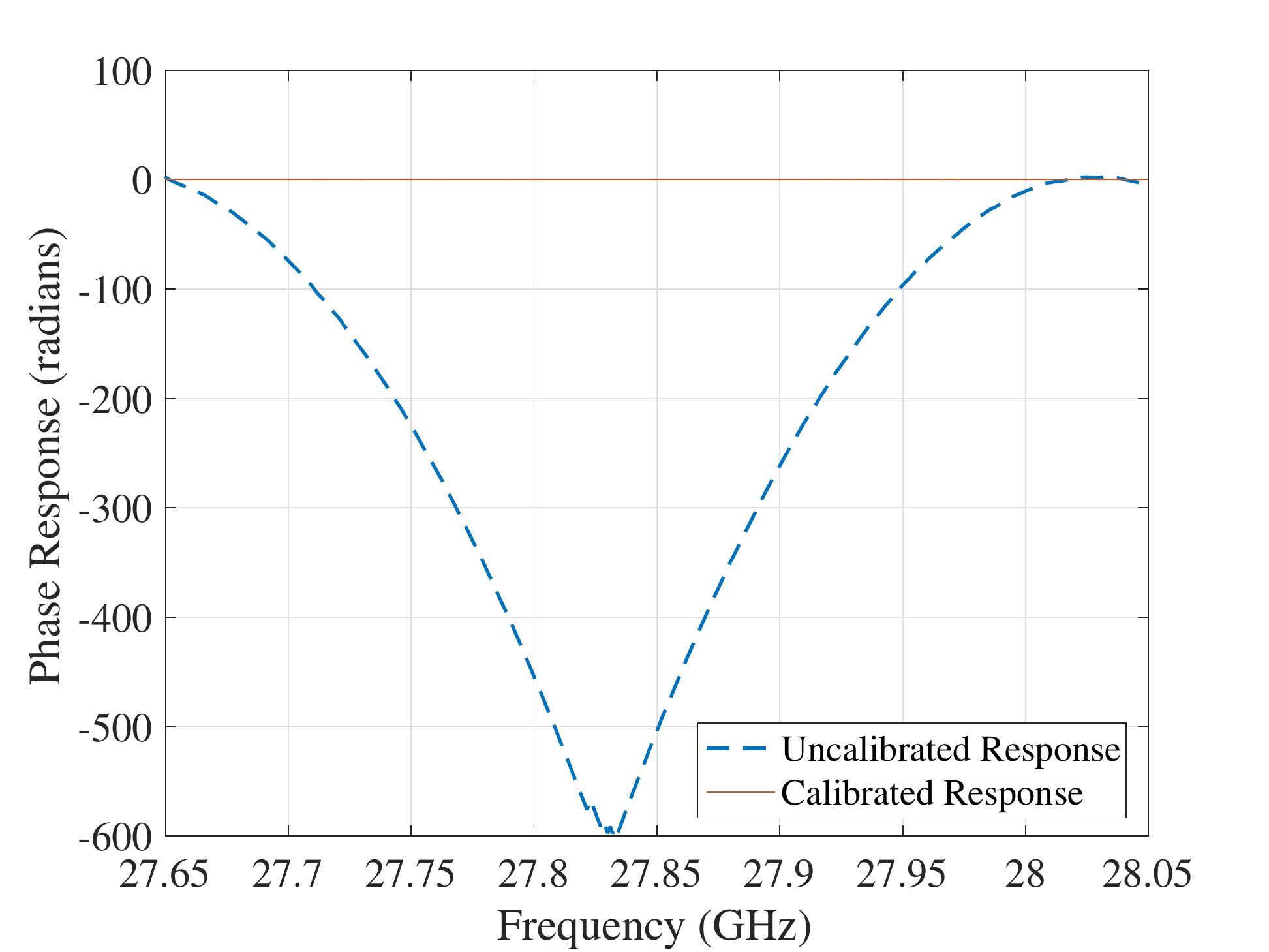}\caption{Calibrated and uncalibrated phase responses of the channel sounder}\label{fig:phase}
\end{figure}

\subsection{Frequency Response}

Since the sounding signal has 400~MHz bandwidth, the frequency response of the channel sounder is also relevant. Figs. \ref{fig:frequency} and \ref{fig:phase} show the amplitude and the phase responses for the channel sounder, respectively. The shown responses are measured baseband to baseband including effects of all the units in the baseband, IF and RF for both TX and RX. The root mean square error (RMSE) for the calibrated amplitude response with respect to the ideal case of unit gain and zero phase is 0.13~dB. Similarly, the RMSE for the calibrated phase response is 0.04 radians (2.3$^\circ$).

\begin{figure}[tbp]
	\centering
	\includegraphics[width=0.6\linewidth]{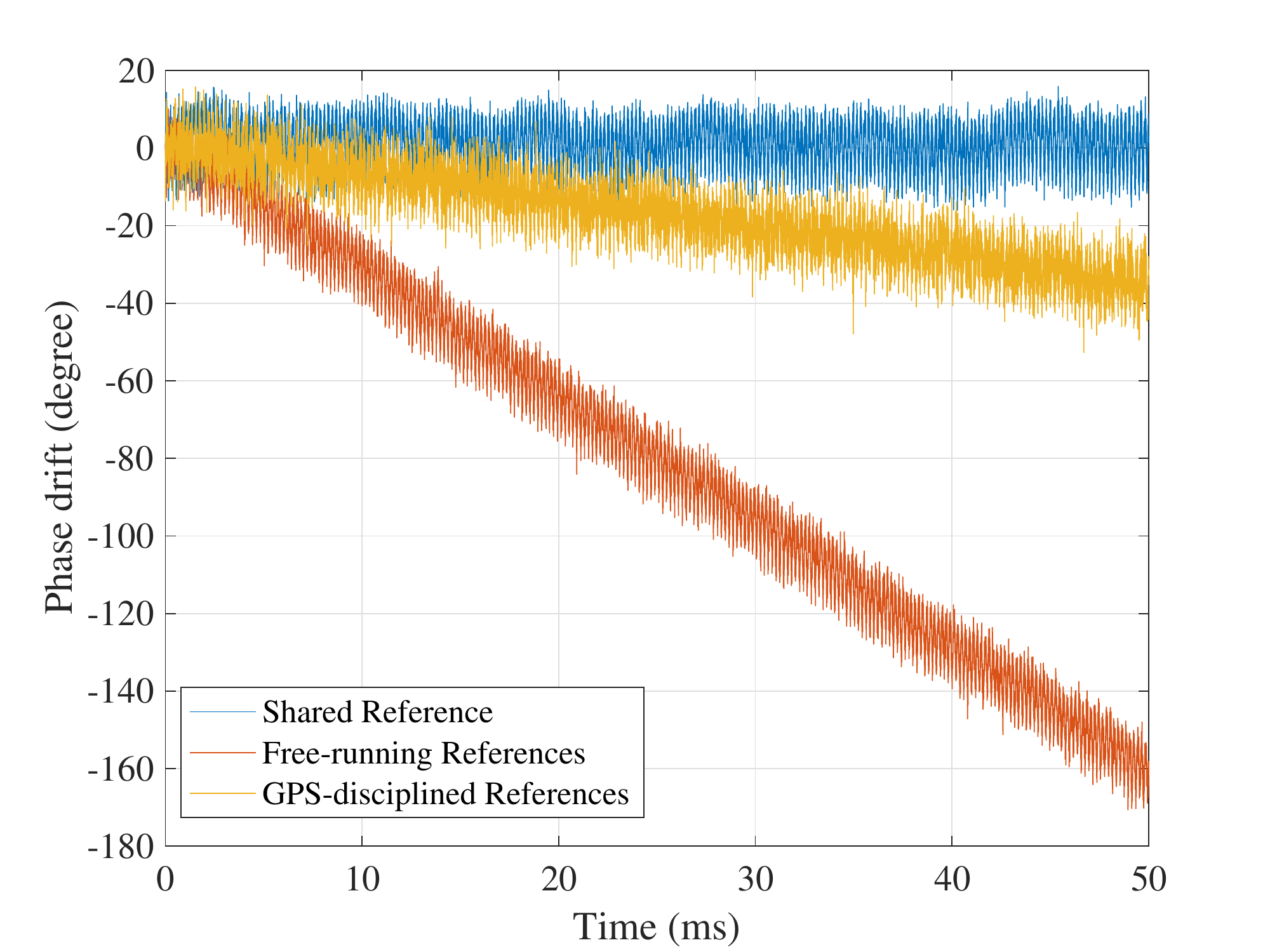}\caption{Phase drift with shared reference, free-running references and GPS-disciplined references}\label{fig:drift}
    \vspace{0.1in}
	\includegraphics[width=0.6\linewidth]{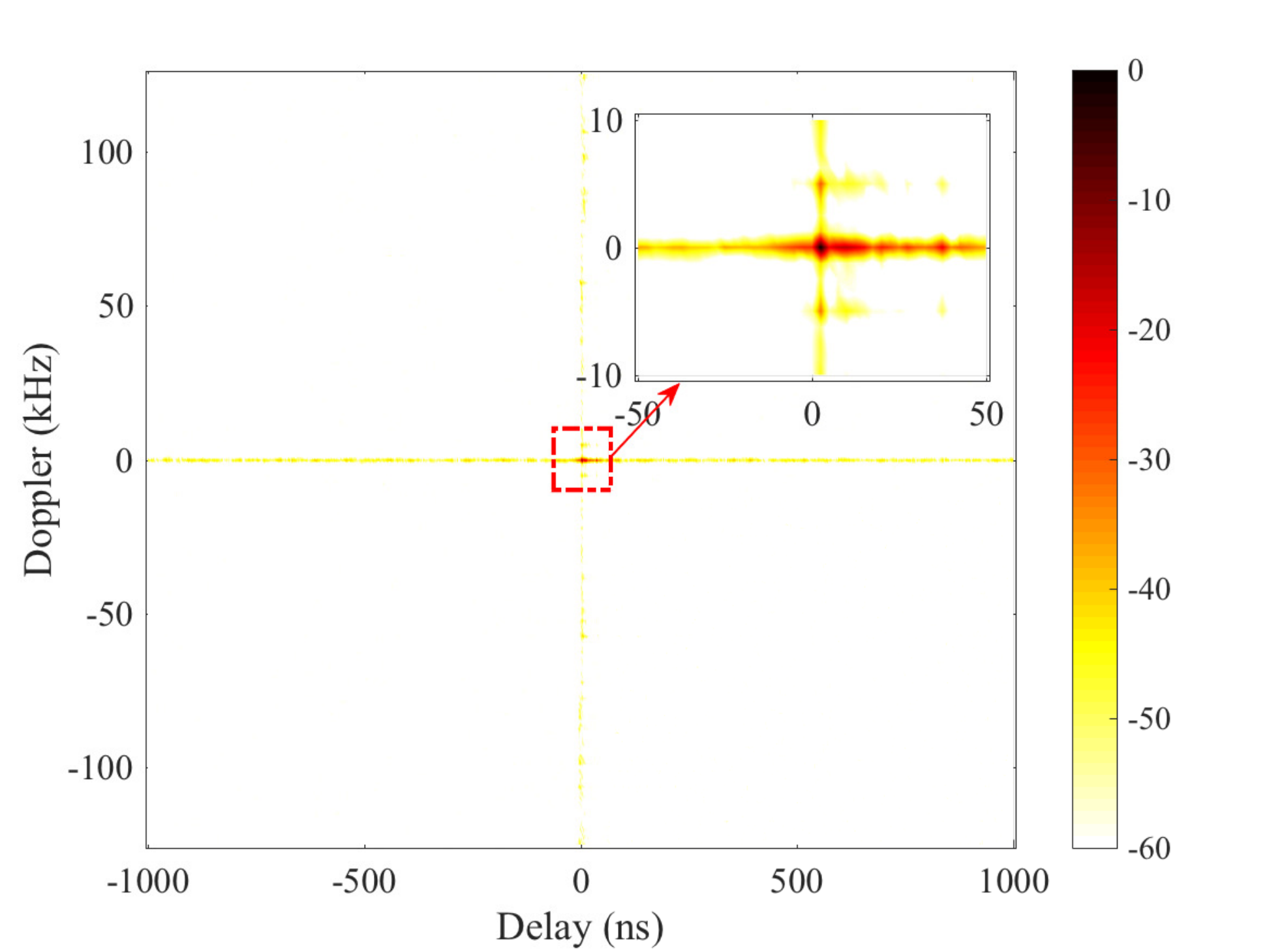}\caption{Normalized delay-Doppler function in dB scale for the measurements with the GPS-disciplined references}\label{fig:delay_doppler}
\end{figure}

\subsection{Phase Stability}\label{sec:phase}

The most important features of the proposed sounder are its short measurement time and phase stability. These two features are required for measurements in dynamic environments. To investigate the phase stability of the system, we run the sounder continuously for \SI{50}{ms} with fixed beams at the TX and the RX.  Fig. \ref{fig:drift} shows the phase drift of the center tone for i) the best-case scenario: TX and RX use a shared reference, ii) the worst-case scenario: they operate with independent free-running Rubidium frequency references, and iii) the typical measurement configuration where the independent Rubidium references disciplined by GPS receivers. With the shared reference there is no phase drift and the standard deviation of the phase is 5.8$^\circ$. Even with the free-running references, the total accumulated phase drift observed during a full-sweep of all (19 by 19) TX and RX beam combinations in azimuth is only 4$^\circ$ over \SI{1.444}{ms}. With the help of GPS-disciplining, this further decreases to less than 1$^\circ$ for the given measurements. During the measurement campaigns, although the performance of the GPS-disciplined case will depend on several factors (i.e. how long the GPS receivers have been connected to GPS satellites, number of active satellites and the quality of the connection), the true phase stability performance will be in between the first two cases. In case of measurements where the GPS might not be available, we also have the option of training one frequency reference with respect to the other prior to measurements to decrease the phase drift for the free-running references. However, this requires a long training duration (12-24~hours), and provide synchronization for only a few hours once the references are separated. 

Fig. \ref{fig:delay_doppler} shows the delay-Doppler function $\chi(\tau,v)$ of the channel sounder for a short distance LOS measurement in a completely static environment. Thanks to the phase stability of the channel sounder, the delay-Doppler function acquired is a good approximation of the ideal case $\chi(\tau,v)=\delta(\tau)\delta(v)$, where $\tau$ is the delay, $v$ is the Doppler shift and the $\delta(\cdot)$ is the Dirac-delta function.

The phase stability is important for determining the achievable averaging (or spreading) gain. Complex averaging achieves noise reduction because the desired signal is added up phase coherently, while noise is added up incoherently. Thus, we can achieve benefits only for averaging durations less than the phase coherence time of the sounder. 

For the admissible measurement time, we have to distinguish between (i) static measurements, and (ii) measurements in dynamic environments. In the former case, the averaging duration {\em within one beam} can be the sounder phase coherence time. This means that up to 10,000 repetitions of the training signal can be made, leading to a spreading gain of 40~dB, and thus extending the measurable path loss to more than 200~dB.
Phase drift between the beams can be estimated easily by simply adopting a switching pattern with repeated
beam pairs. The estimation is further simplified by the fact that the phase drift is essentially
linear even over the fairly long time of 50~ms \cite{Kristem_2017_Channel}. In the case of dynamic environments, a complete MIMO snapshot has to be finished within a time that is the shorter of the sounder phase coherence time and the channel coherence time. In a typical configuration, we employ 10 repetitions for each beam pair to improve the SNR by averaging resulting a sweep time of \SI{14.44}{ms}. In this case, the accumulated drift is 40$^\circ$ in the worst case.

\begin{figure}\centering
	\includegraphics[width=0.55\linewidth, viewport=38 180 550 600, clip=true]{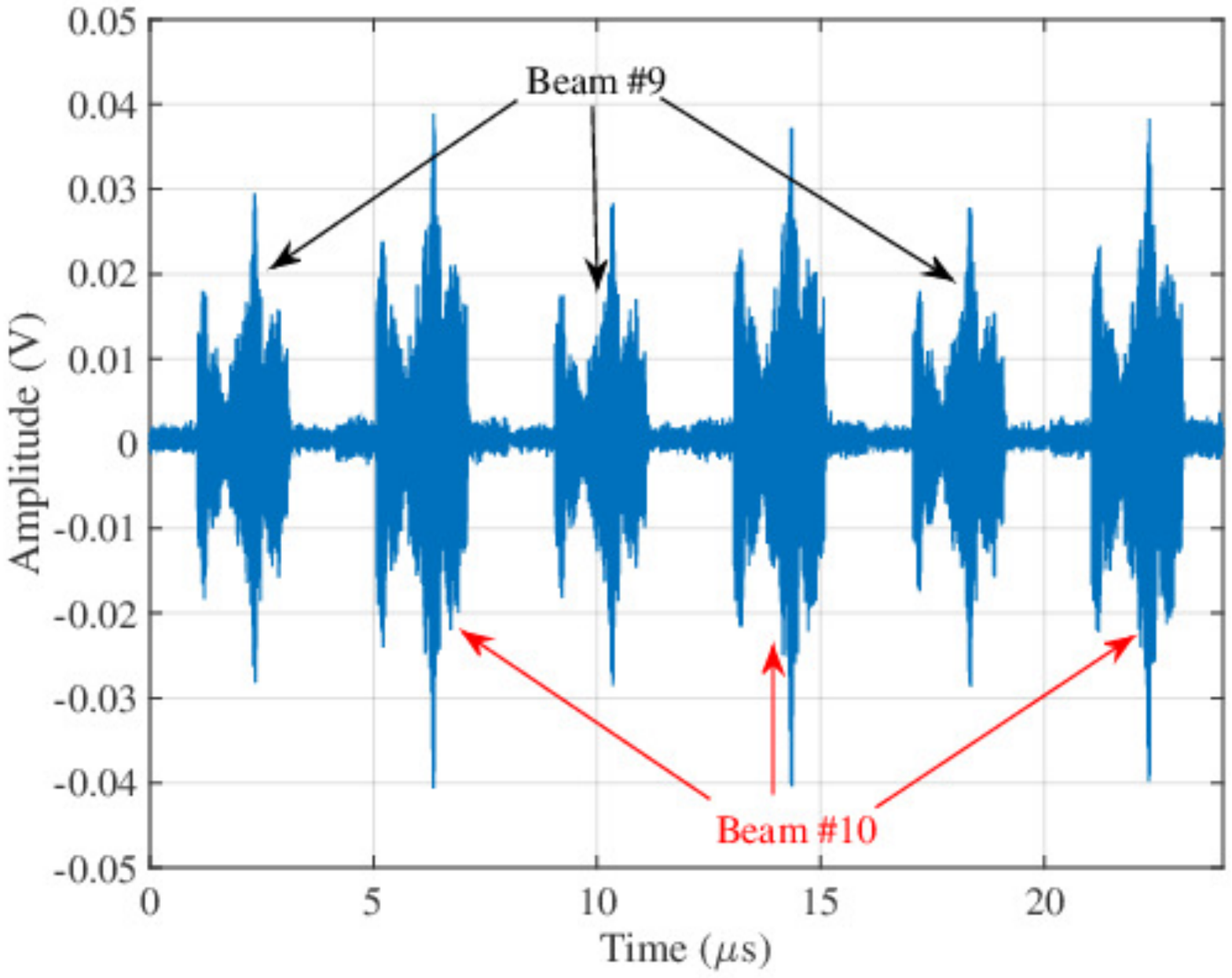}\caption{Received waveform while switching between 2 beams; Beam $\#$9 and Beam $\#$10 with azimuth directions of $-5^\circ$ and $0^\circ$ }\label{fig:beam_switch}
    \vspace{0.1in}
	\includegraphics[width=0.55\linewidth, viewport=38 180 550 600, clip=true]{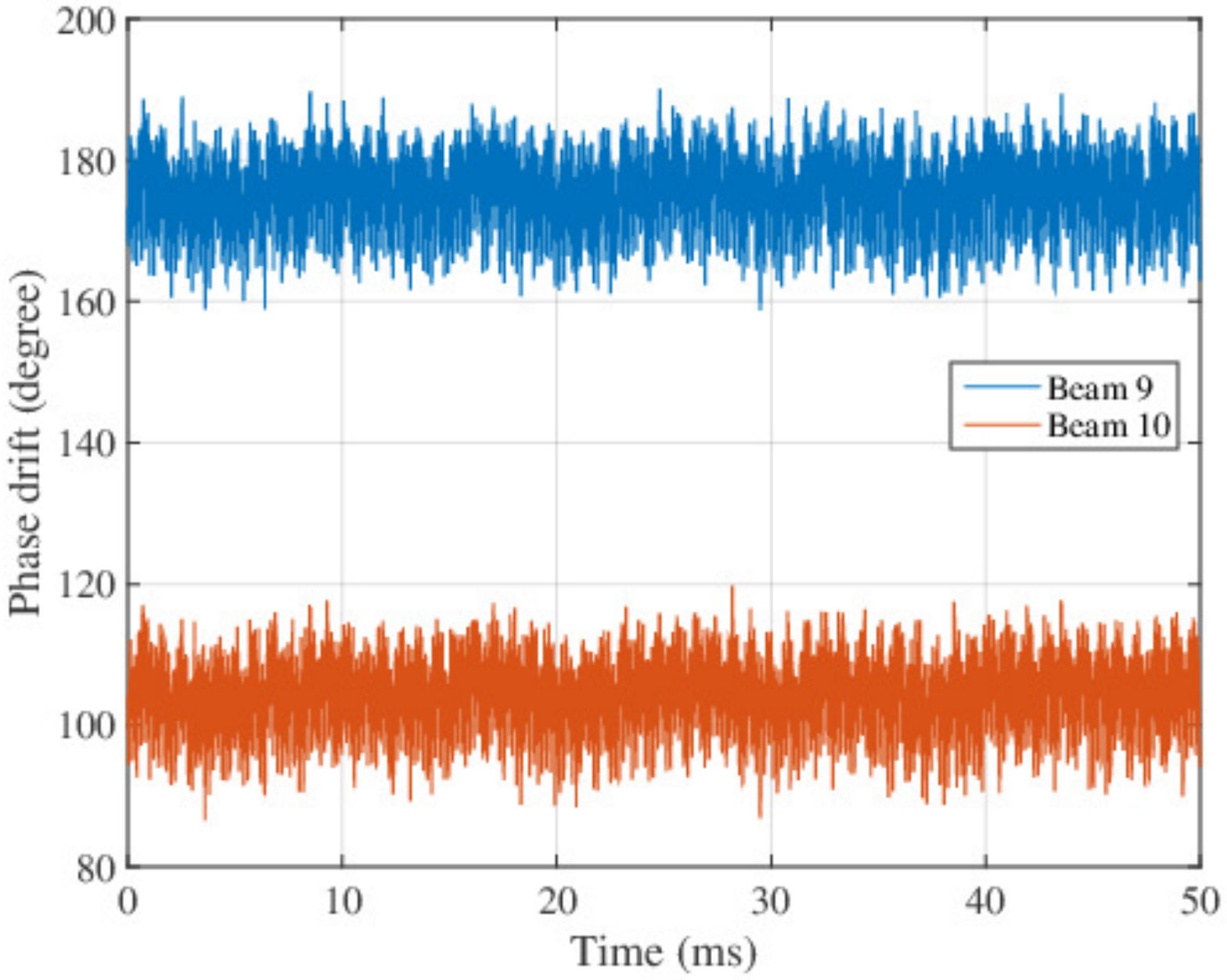}\caption{Phase drift for Beam $\#$9 and Beam $\#$10 measured while switching between 2 beams} \label{fig:beam_phases}
\end{figure}

The repeatability of the phase is as important as its stability when investigating temporal changes (i.e. Doppler spectrum) in dynamic channels. To ensure that the calibration measurements are valid, it is crucial that the relative phase between 2 different beams stays constant at all times, even after a complete restart of the sounder. To test this, we recorded the received waveform, while switching back and forth between two beam settings as seen in Fig. \ref{fig:beam_switch}. Fig. \ref{fig:beam_phases} shows the phase of the center tone for the two beams. The phase offset between two beams does not change significantly over 50~ms. Furthermore, this phase offset does not change even after a complete power cycle of the sounder. Both in the frequency synthesizers and in the BF-RFUs, we have phase locked loops (PLLs) to derive the carrier frequencies from 10 MHz references. Every time they lock, they do so with a random phase, however, the effect of the random phase is the same for all beam pairs and has no effect on the {\em relative} phase offset between different beams, which is the relevant quantity to the post-processing.

\begin{figure}\centering
	\includegraphics[width=0.6\linewidth]{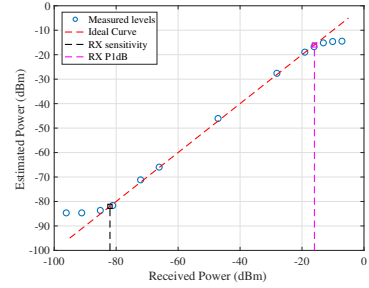}\caption{Dynamic range of the RX, black dashed line indicates the RX sensitivity level, magenta line indicates the RX 1~dB compression point} \label{fig:rx_range} 
\end{figure}

\subsection{Dynamic Range and Measurable Path Loss}\label{sec:dynamic range}
The estimated noise figure for the RX is 5~dB. With 400~MHz measurement bandwidth, the RX sensitivity is given by
\begin{equation}
 \resizebox{.9 \linewidth}{!}{
$RX_{\rm sens}= -174~{\rm dBm/Hz}+ 5~{\rm dB} + 10*log_{10}(400e6)~{\rm Hz} = -83~dBm$
}
\label{eq:rxsens}
\end{equation}

\noindent Fig. \ref{fig:rx_range} shows the measured dynamic range of the RX. The received power on the x-axis is the calculated power given the transmitting power and TX beam forming gain\footnote{Note that since the up-conversion and phased-arrays can not be separated, these measurement are performed over the air in an anechoic chamber. Different power levels are achieved by changing the TX power and TX-RX beam combinations.}. The power on the y-axis is the estimated power from the recorded waveform averaged over the whole band. In parallel to the estimated RX sensitivity, noise power distorts the estimated power if the input power is below $-83$~dBm. In addition, the RX starts to saturate at $-6$~dBm. Hence, the channel sounders dynamic range is 77~dB.

Note that the sensitivity given in Eq. (\ref{eq:rxsens}) does not include the RX array gain $G_{\rm RX}(dB)$; the equivalent isotropic sensitivity (EIS) is $-102$~dBm given by $EIS(dB)=RX_{\rm sens}-G_{\rm RX}(dB)$ \cite{zhang2017antenna}. Combined with the maximum 57~dBm EIRP, the instantaneous measurable path loss for the channel sounder is 159~dB. In comparison, the state-of-the-art rotating horn antenna channel sounder presented in \cite{MacCartney_2017_flexible} can measure path loss up to 185 dB. However, this channel sounder's operating principle is based on sliding correlator which requires a much longer measurement duration. When using the same acquisition time of 655.04~ms, we could with our sounder repeat the same measurement 327500 times (assuming sufficient phase stability, as discussed above) which would increase the measurable path loss by 55~dB to a total of 214~dB via complex RX waveform averaging. To provide another comparison, the state-of-the-art channel sounder presented in \cite{sun2017design} has 136~dB measurable path loss excluding the averaging and processing gains.

\begin{figure}[tbp]
	\centering\includegraphics[width=0.6\linewidth,viewport=38 180 550 600, clip=true]{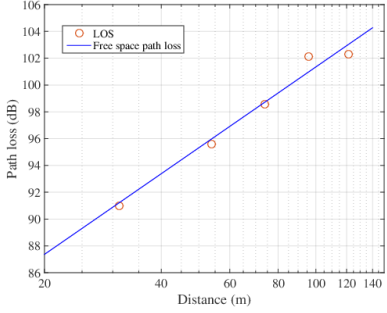}\caption{Line-of-sight path loss versus TX-RX distance}\label{fig:los_pl}
\end{figure}

\subsection{Path Loss Verification}

Fig. \ref{fig:los_pl} shows the measured path loss for line-of-sight (LOS) measurements conducted in an open area when the TX and the RX were placed on scissor lifts at the height of 5~m. The measurements were performed for distances ranging from 30~m to 122~m. The path loss exponents were estimated 1.997 and 2.051 for close-in and alpha-beta-gamma models, respectively. For both models, the observed path loss exponents were almost equal to the free-space path loss exponent of 2. The absolute path loss measured in the anechoic chamber at the distance of \SI{6}{m} is \SI{18.25}{dB} when the bore-sight of the center beams for the TX and RX are aligned. After taking the gains of \SI{12}{dB} for amplifiers and the \SI{17}{dBi} for the beamforming (for both the TX and the RX) into account, this corresponds to \SI{76.25}{dB} over the air path loss showing a good agreement with the  free space path loss for the same distance, which is \SI{76.9}{dB}.

\section{Post Processing} \label{sec_post}
In the following, we describe the post processing of the measured data using Fourier-resolution techniques. 

\subsection{Power Delay Profile and Power Angular Delay Profile}

The directional power delay profile (PDP) for the TX and RX beams with the azimuth angles  $\phi_\text{TX}$  and  $\phi_\text{RX}$, respectively, is estimated as
\begin{equation}  
 \resizebox{.9 \linewidth}{!}{
$  PDP(\phi_\text{TX},\phi_\text{RX},\tau) = \bigg\vert \mathcal{F}^{-1} \left\{W\left(\vec{f}\right) \cdot H_{ \phi_\text{TX},\phi_\text{RX}}\left(\vec{f} \right) ./ H_\text{cal}\left(\vec{f} \right) \right\} \bigg\vert ^2$ }
\end{equation}

\noindent where $\phi_\text{TX/RX}\in[-45,45]$, $\mathcal{F}^{-1}$ denotes inverse Fourier transform, $H_{i,j}(\vec{f})$ and $H_\text{cal}(\vec{f})$ are the frequency responses for $i$-th TX and $j$-th RX beam and the calibration response respectively; $W(\vec{f})$ is the Hanning Window, $\vec{f}$ is the vector of frequency tones, and $./$ is element-wise division. 

Since all beam pairs are measured without a significant phase drift or trigger jitter, all directional PDPs are already aligned in the delay domain and require no further correction. Consequently, the corresponding PDP can be obtained similarly to the virtual horn measurements with a shared reference even though the sounder does not have a shared reference. There are different methods discussed in \cite{hur_synchronous_2014,Haneda_2016_omni} for synthesizing omni-directional PDP from directional measurements. We use the approach from \cite{hur_synchronous_2014} to present sample results here. Hence, the omnidirectional PDP is given by
\begin{equation}
  PDP(\tau) = {\displaystyle \max_{\phi_\text{TX}} \max_{\phi_{RX}}  PDP(\phi_\text{TX},\phi_\text{RX},\tau)} \label{eq:pdp}
\end{equation}

Furthermore, the power angular-delay profiles for RX and TX which are calculated as follows. 
\begin{equation} 
\begin{aligned}  
  PADP_\text{RX}(\phi_\text{RX},\tau) &= {\displaystyle \max_{\phi_\text{TX}} PDP(\phi_\text{TX},\phi_\text{RX},\tau)} \\
  PADP_\text{TX}(\phi_\text{TX},\tau) &= {\displaystyle \max_{\phi_\text{RX}} PDP(\phi_\text{TX},\phi_\text{RX},\tau)}
 \end{aligned} 
\end{equation}

Then the angular power spectrum (PAS) can be simply calculated as:
\begin{equation}
 {\displaystyle  PAS(\phi_\text{TX},\phi_\text{RX}) = \sum_{\tau} PDP(\phi_\text{TX},\phi_\text{RX},\tau)}
\end{equation}

\begin{algorithm}[tbp]
  \caption{\textbf{Multi-path detection:}} \label{alg_detect}
\begin{algorithmic}[1]  \footnotesize
  \Procedure{detect MPC}{$PDP(\phi_\text{TX},\phi_\text{RX},\tau)$,$N_{th}$}
    \State Initialize MPClist
    \State Perform 3-D peak search on $PDP(\cdot)$
    \State Store peaks $p(\tau,\pmb{\phi})$, $\pmb{\phi}=[\phi_\text{TX},\phi_\text{RX}]$
    \ForAll{$\tau$}
      \State $\tau_n \gets \tau$ 
      \If{$p(\tau_n,\pmb{\phi})$ is not empty}
        \State $p_\text{max} \gets \max\limits_{\pmb{\phi}} p(\tau_n,\pmb{\phi})$
        \State $\pmb{\phi}_\text{max} \gets arg\max\limits_{\pmb{\phi}} p(\tau_n,\pmb{\phi})$
        \ForAll{$p(\tau_n,\pmb{\phi})$}
          \If{$p(\tau_n,\pmb{\phi}) > p_\text{max}/10$ }
            \State Add $p(\tau_n,\pmb{\phi})$ to the MPClist      
          \ElsIf{$p(\tau_n,\pmb{\phi}) > p_\text{max}/100$ }
              \If{$(\phi_\text{TX,max}\neq\phi_\text{TX})$\&$(\phi_\text{RX,max}\neq\phi_\text{RX}$)}
                \State Add $p(\tau_n,\pmb{\phi})$ to the MPClist 
              \EndIf
          \EndIf
        \EndFor
      \EndIf
    \EndFor
  \EndProcedure
\end{algorithmic}
\end{algorithm}

\subsection{Multi-path Extraction}

The MPC extraction from the measured directional PDPs is described in Algorithm \ref{alg_detect}. For extracting the MPCs in the delay domain, we consider separately each resolvable delay bin. Within each bin, obtaining the MPC directions cannot simply assign an MPC to each beam with significant power; rather we have to take the side-lobes of the beams into account. As seen in Fig. \ref{fig:pattern} and \ref{fig:beams}, in all cases, the main lobe is at least 10 dB stronger than the side-lobes. We first perform a peak search for all peaks in the TX/RX beam domain with power that is higher than a given threshold $N_{\rm th}$, which is set as average noise power plus \SI{6}{dB} in this work. Subsequently, we filter out the ghost MPCs due to side-lobes, a ghost MPC must have the same delay with a stronger valid MPC since (due to our Hanning filtering), no significant side-lobes exist in the delay domain. The strongest peak detected at every delay bin is always accepted as a MPC. Since the beam-width is wider than the beam steering steps, a MPC will likely be received in more than a single beam. However, thanks to the concave shape of the beam pattern within the main beam each MPC will correspond to a single peak in the PADP. A peak that has the same delay and a power within \SI{10}{dB} of the strongest MPC is also accepted as a valid MPC, since it cannot represent a side-lobe. Finally, we accept the peaks that have powers within \SI{20}{dB} of the strongest MPC and both the direction of departure (DoD) and the direction of arrival (DoA) are different than the strongest MPC in the same delay bin, since ghost MPCs fulfilling that angle condition would be suppressed by twice the side-lobe suppression (once at TX, once at RX), and thus be suppressed by more than 20 dB. Note that with this extraction algorithm, the dynamic range {\em per delay bin} is limited to 20 dB (even for very high SNR), though for different delays, MPCs within a larger dynamic range can be detected.

\begin{figure}[tbp]
	\centering\includegraphics[width=0.65\linewidth]{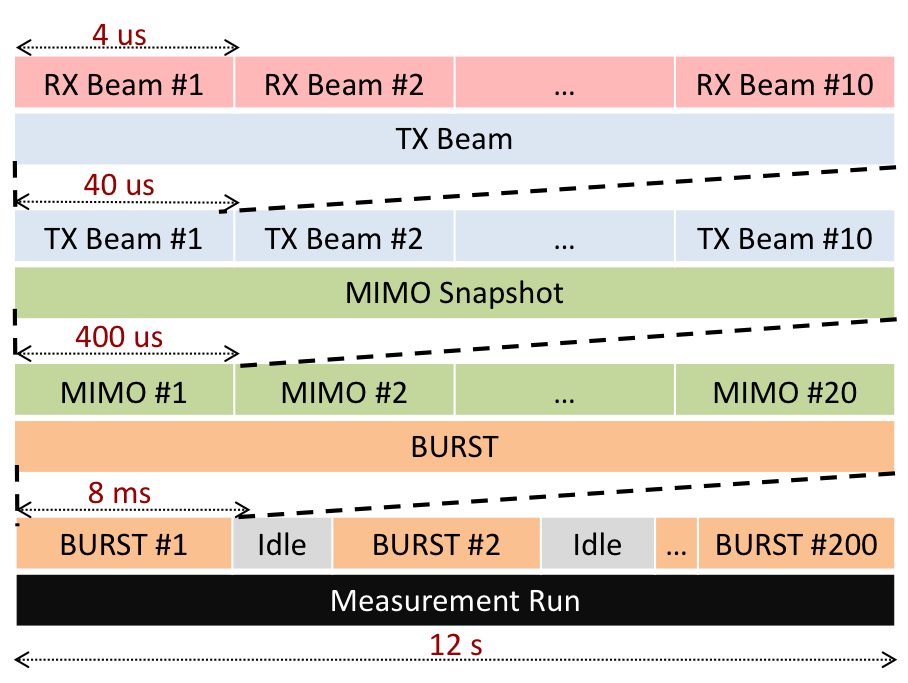}\caption{Time line for the channel sounder operation}\label{fig:switching}
\end{figure}

\section{Sample Measurements}\label{sec_meas}

To demonstrate the capability of the sounder, this section will discuss sample results from a first-of-its-kind directionally resolved dynamic measurement campaign \cite{bas_2017_dynamic}. Figure \ref{fig:switching} shows the time-line for the sounder operation for this measurement campaign.  Each SISO measurement (sounding of a TX and a RX beam) takes \SI{4}{\mu s} consisting of \SI{2}{\mu s} sounding waveform and \SI{2}{\mu s} guard time for electronic beam-switching. Although the channel sounder is capable of beam-steering with $5^\circ$ steps covering $\pm 45^\circ$, to decrease the measurement time we used every other beam resulting in $10^\circ$ angular resolution. Consequently, both TX and RX perform $90^\circ$ azimuth sweeps measuring 100 beam pairs in only $400 \mu s$. A single sweep of all possible combinations of TX and RX beams is called a MIMO snapshot. In a burst, 20 MIMO snapshots were measured without any idle time in between. This allows us to estimate Doppler shifts up to $\SI{\pm 1.25}{kHz}$, which corresponds to a maximum relative speed of \SI{48}{kph} at \SI{28}{GHz}, which is larger than the maximum permissible speed on the measured street. Furthermore, these \SI{8}{ms} bursts of MIMO measurements followed by \SI{52}{ms} of idle time were repeated 200 times with a period of \SI{60}{ms} to track the evolution of channel parameters as they change due to moving objects in the environment. This configuration provides us a unique capability of performing double-directional measurements under dynamic conditions and investigating the effects of moving objects on the angular channel characteristics and Doppler spectrum. During the measurements, the TX was placed on a scissor lift and the antenna array was at a height of 3.5~m similar to a micro-cell while the RX height was 1.8~m emulating a user equipment. 

\begin{figure}[tbp]
	\centering\includegraphics[width=0.65\linewidth]{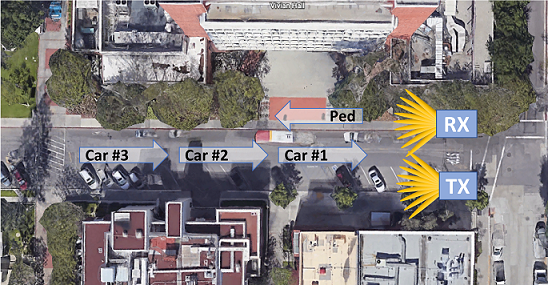}\caption{The environment for the dynamic measurements Case 1 with moving scatterers, three cars moving towards to TX-RX, a pedestrian walks away from the TX-RX }\label{fig:rx2_run8}
	    \vspace{0.1in}
	\centering\includegraphics[width=0.65\linewidth, viewport=10 165 620 620, clip=true]{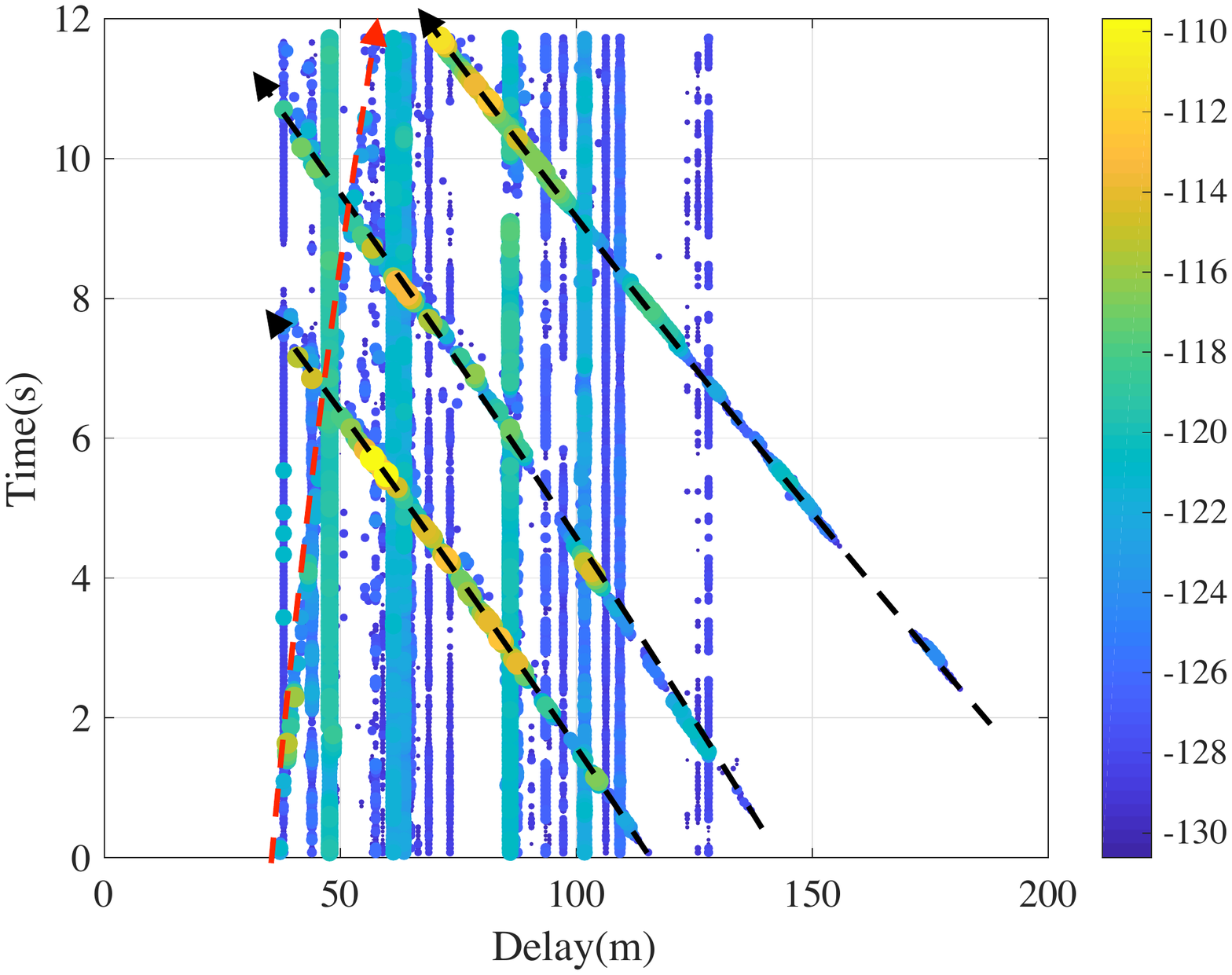}\caption{PDP(dB) vs time for Case 1, 3 moving cars are marked with black arrows and the pedestrian is marked with red arrow} \label{fig:rx2run8_pdp}
	    \vspace{0.1in}
	\centering\includegraphics[width=0.75\linewidth]{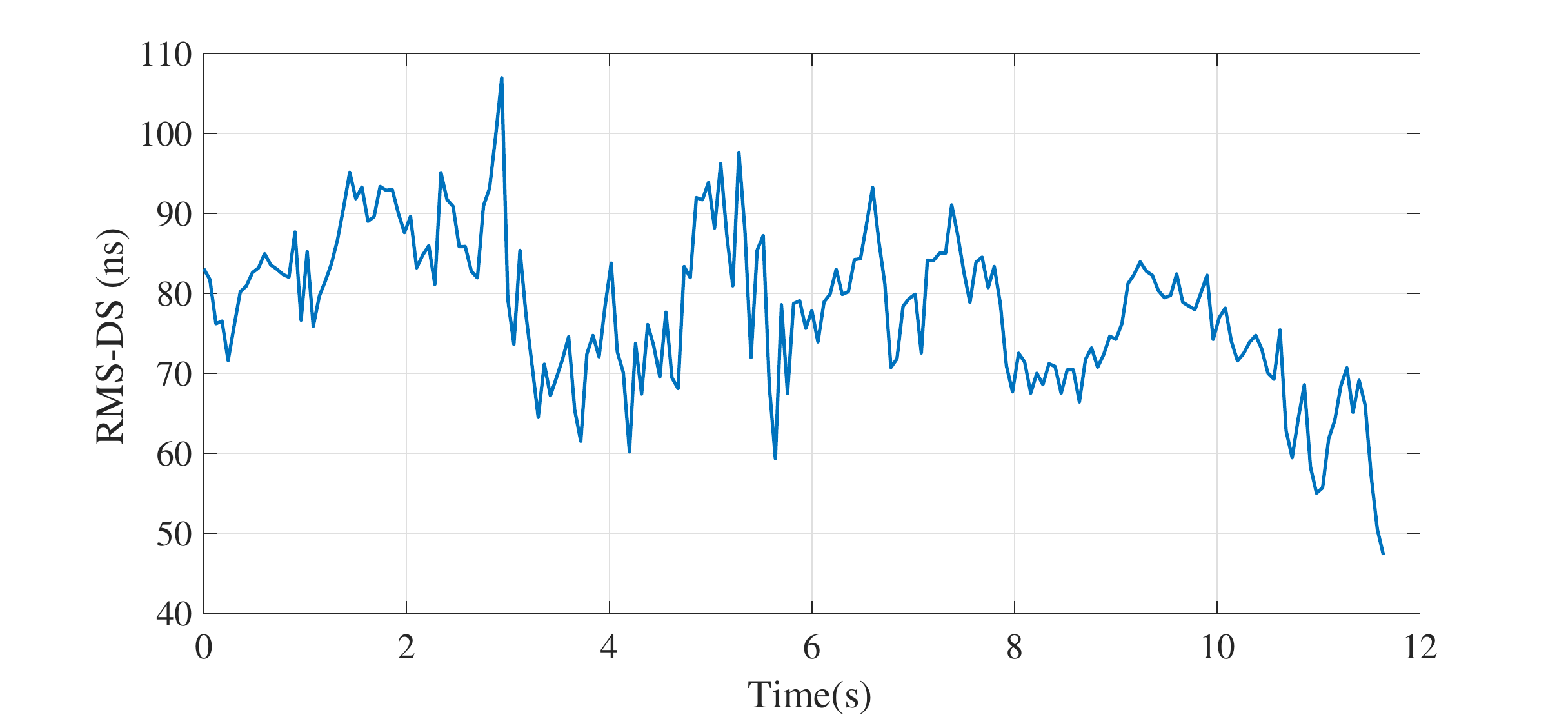}\caption{Time varying root mean square delay spread for Case 1} \label{fig:rx2run8_rmsds}
\end{figure}

\subsection{Case 1: Moving Scatterers}

Fig. \ref{fig:rx2_run8} shows the details about the measurement snapshot presented in this section. During these measurements both TX and RX are placed facing the same direction so that they are out of each other's visible azimuth range. We can thus observe moving scatterers without a dominant LOS. During \SI{12}{s}, we observe 4 moving objects; three cars moving towards the TX and RX and a pedestrian walking away. The tracked MPCs corresponding to these four objects are marked on the PDP in Fig. \ref{fig:rx2run8_pdp}. Fig. \ref{fig:rx2run8_rmsds} shows the root mean square delay spread (RMS-DS) varying within range from \SI{50}{ns} to \SI{100}{ns} through \SI{12}{s} of measurements.

\begin{figure}[tbp]    
	\centering\includegraphics[width=0.75\linewidth]{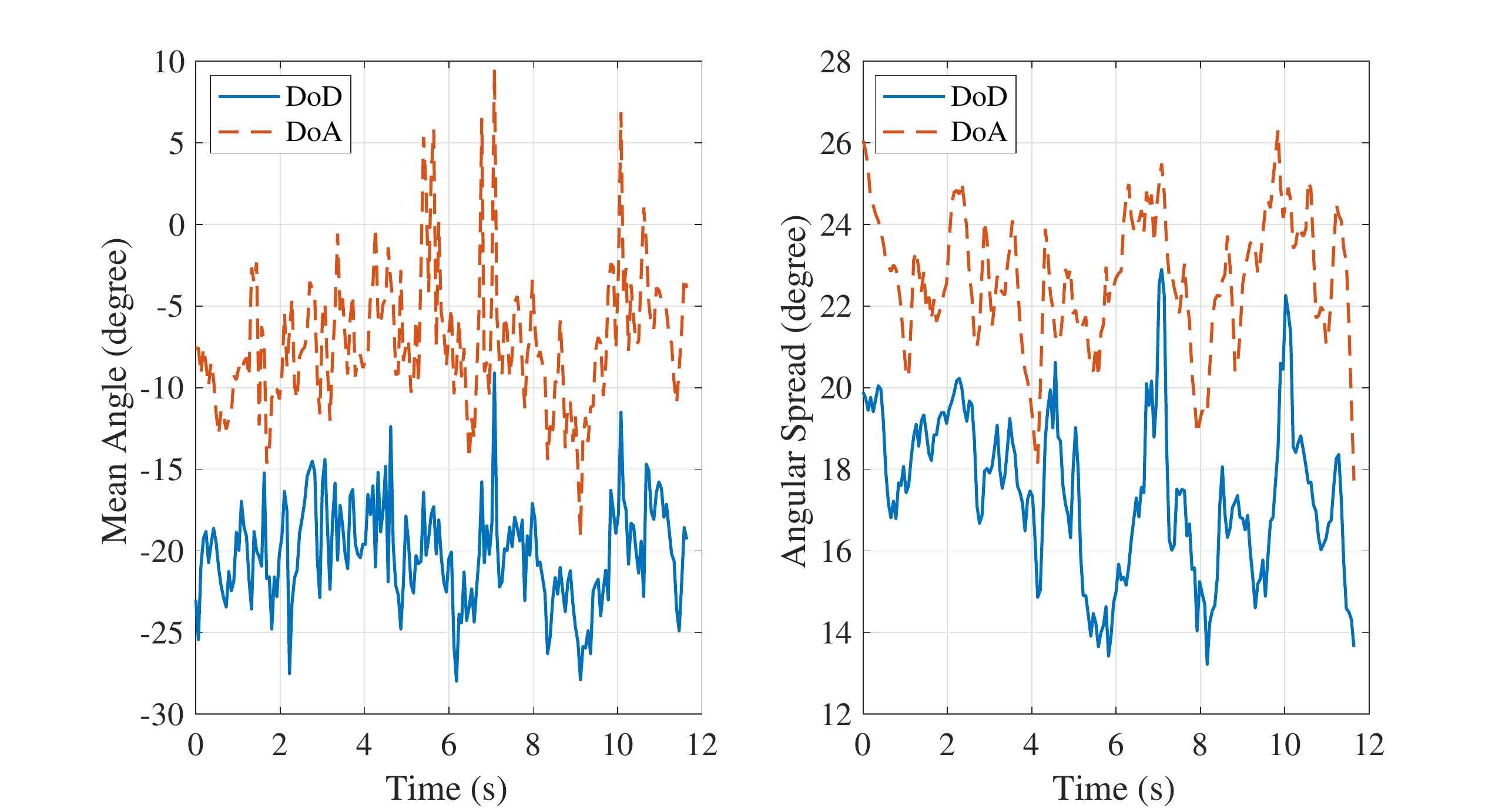}\caption{Mean angle and angular spreads vs time for Case 1}\label{fig:rx2run8_ang}
	\vspace{0.1in}
	\centering\includegraphics[width=0.75\linewidth]{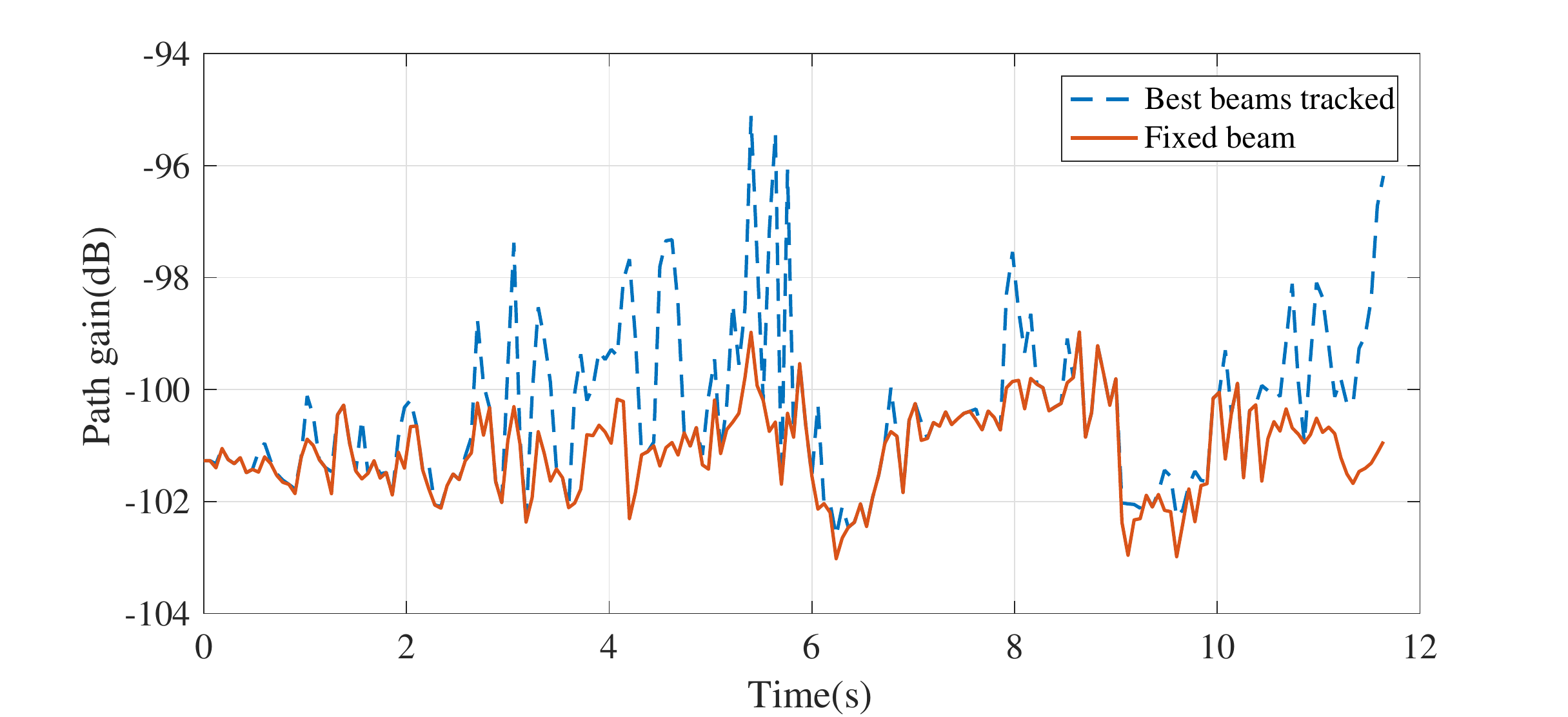}\caption{Path gains vs time: i)RX and TX tracks the best beam pair,  ii) stays on the which best on the average for Case 1}\label{fig:beam_gains}
    \vspace{0.1in}
        \centering\includegraphics[width=0.75\linewidth]{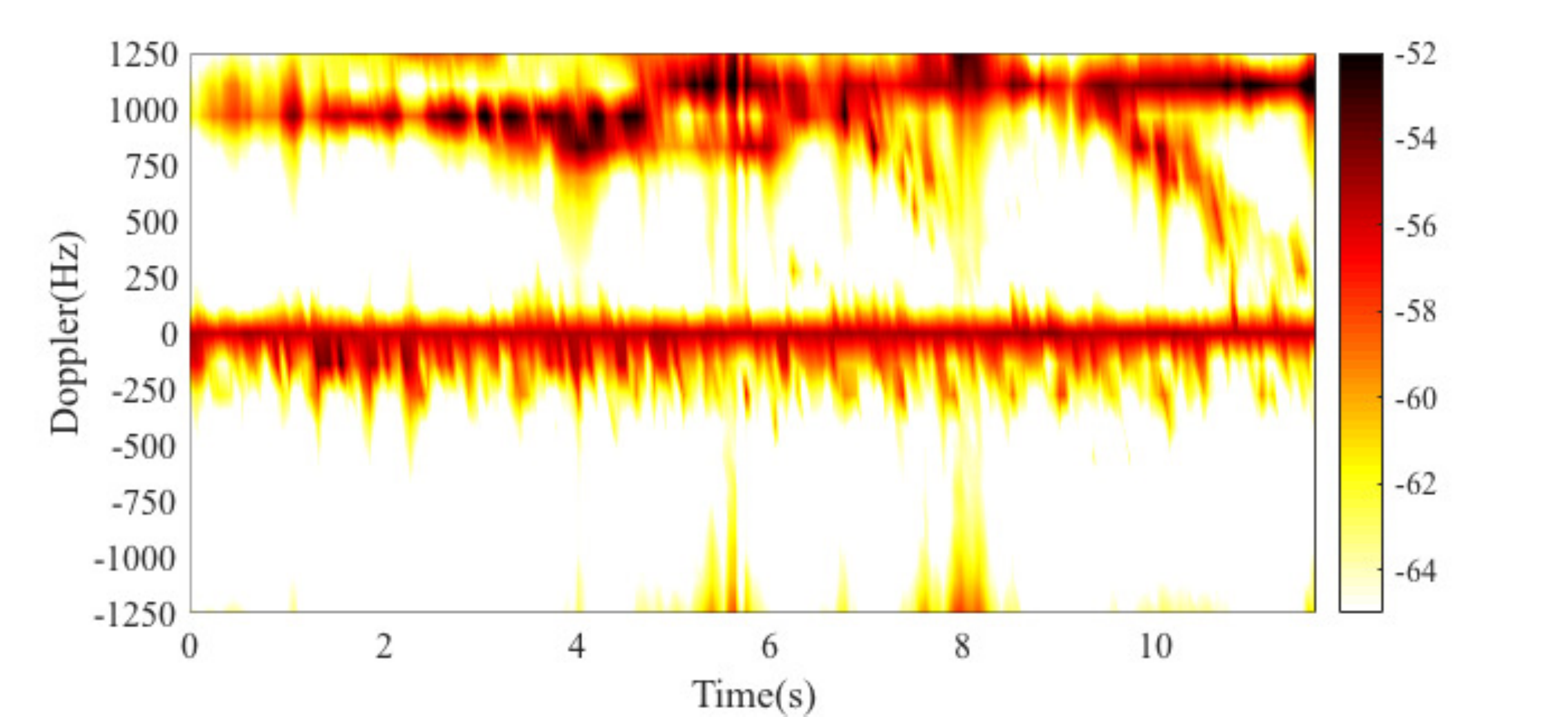}\caption{Doppler spectrum (dB) vs time for Case 1}\label{fig:dopp_time}
    \vspace{0.1in}
        \centering\includegraphics[width=0.75\linewidth]{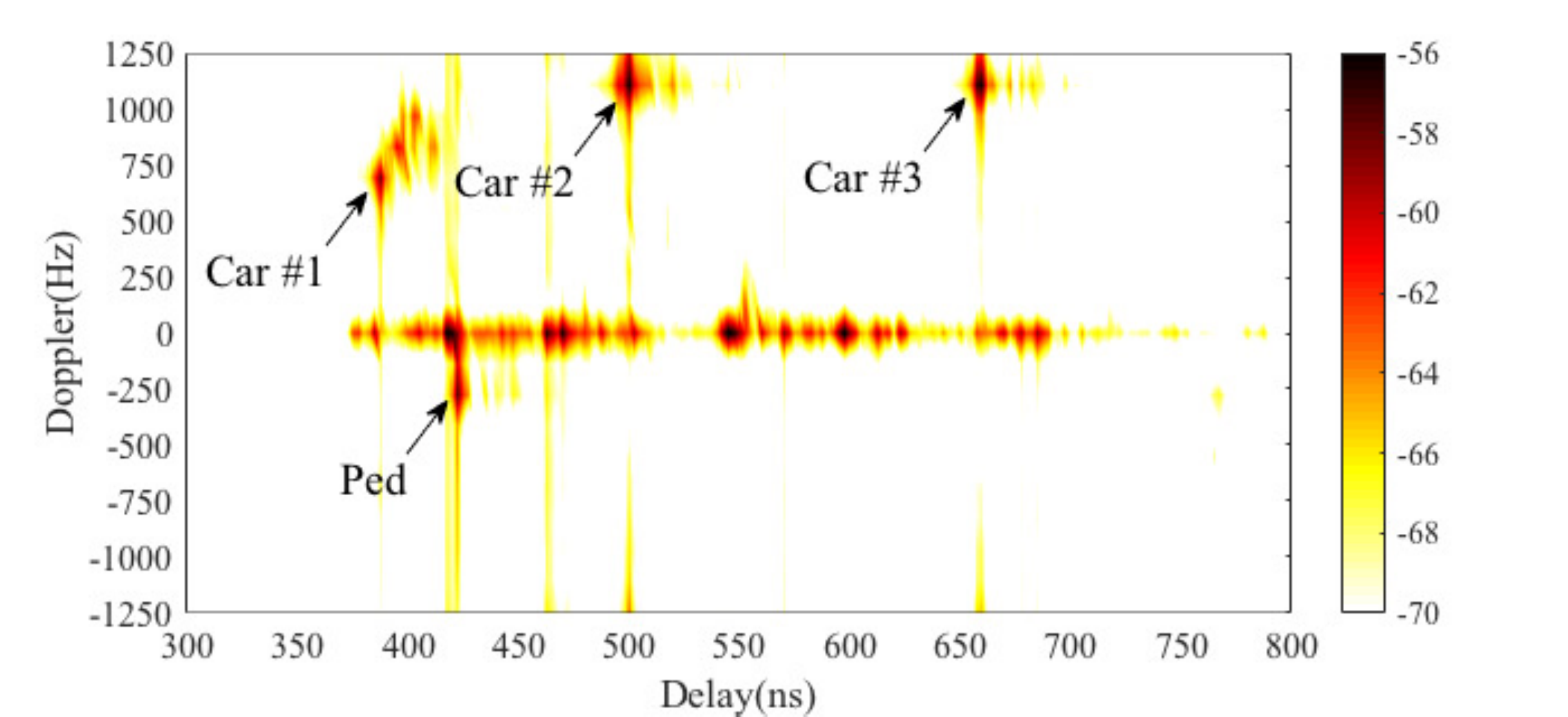}\caption{Doppler spectrum (dB) vs delay at $t=7.38$~s  for Case 1 }\label{fig:dopp_delay}
\end{figure}

Thanks to the capability of performing directional measurements in dynamic environments, we can also study the evolution of the angular channel statistics. Due to fast moving objects in the environment, and their interactions with each other, the angular spectra for the TX and the RX change quickly. As seen in Fig. \ref{fig:rx2run8_ang}, the mean DoA angle varies from $-15^\circ$ to $10^\circ$ and the angular spread takes values in the range of $18^\circ$ to $26^\circ$. Similarly, the DoD varies between $-27^\circ$ and $-10^\circ$ with angular spreads from $14^\circ$ to $23^\circ$.

While designing a mm-wave system with adaptive beam-forming it is crucial to understand the requirements on the beam-switching, i.e., how frequently we have to search for the best beam and when the beam should be switched. Hence, the measurements which can capture the path gain simultaneously for different TX and RX beam directions are needed. Fig. \ref{fig:beam_gains} shows the path gains for the fixed beam with highest average power and if we track the best beam at all times. At times, the instantaneous best beam surpasses the fixed beam by more than 8 dB as the channel changes due to moving objects.

The phase stability of the system allows us to estimate the Doppler spectrum and its temporal evolution. The 20 fast repetitions within each burst are used to estimate Doppler shifts for each time instance. Fig. \ref{fig:dopp_time} shows the Doppler spectrum over time. Since all three cars have relatively similar speeds, they create a similar Doppler shift of 1kHz-1.25 kHz, but they can be distinguished through their different delays, see the spreading function in Fig. \ref{fig:dopp_delay}. Additionally, we observe a small negative Doppler shift due to the pedestrian walking away.

\begin{figure}[tbp]
	\centering\includegraphics[width=0.65\linewidth]{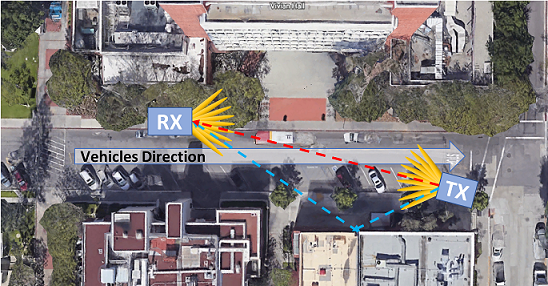}\caption{The environment for the dynamic measurements Case 2 with blocking objects. Two dominant paths observed in the idle channel: i) LOS: red dashed line, ii) the reflection : blue dashed line}\label{fig:dynamic_2}
	\vspace{0.1in}
	\centering\includegraphics[width=0.65\linewidth]{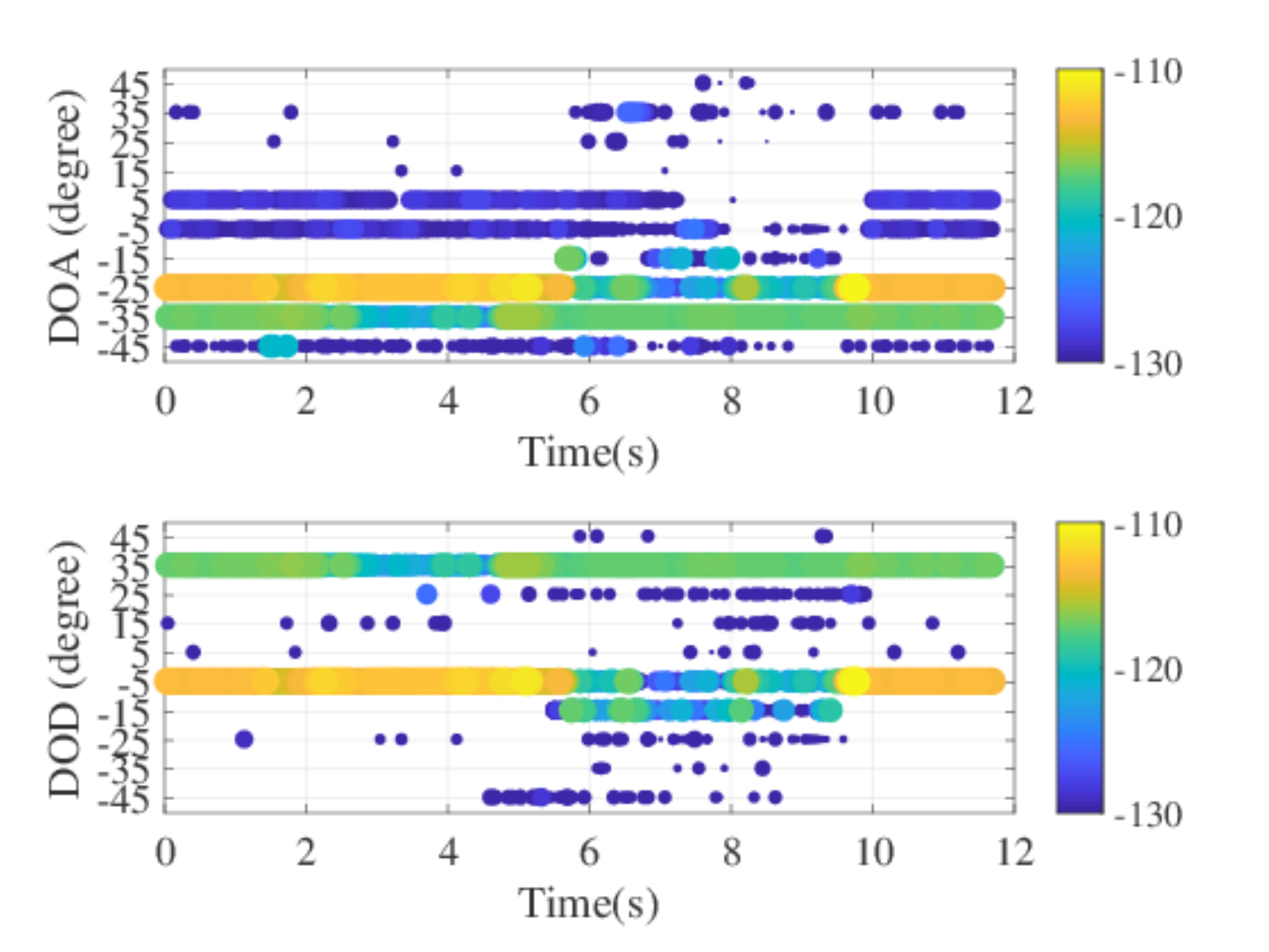}\caption{PAS(dB) for the TX and the RX, the reflection (i.e., $[\phi_{\rm TX},\phi_{\rm RX}]=[35,-35]$) is blocked between 3~s and 4.8~s and LOS (i.e., $[\phi_{\rm TX},\phi_{\rm RX}]=[-5,-25]$) is blocked between 5.5~s and 9~s.} \label{fig:case2_PAS}
\end{figure}

\subsection{Case 2: Blocking Objects}
In this scenario, we observe two main MPCs while the channel is idle; the LOS path and a reflection as shown in Fig. \ref{fig:dynamic_2}. The LOS path has the directions $[\phi_{\rm TX},\phi_{\rm RX}]=[-5,-25]$ and the reflection $[\phi_{\rm TX},\phi_{\rm RX}]=[35,-35]$.

The actions during this measurement can be listed as;
\begin{itemize}
	\item The measurements start with an idle channel,
	\item a truck enters to the street moving left to right,
	\item the truck blocks the reflection between $t=3$~s and $t=4.8$~s,
	\item the truck blocks the LOS path between $t=5.5$~s and $t=9$~s,
    \item The measurements end with an idle channel,
\end{itemize}

\begin{figure}
	\centering\includegraphics[width=0.75\linewidth]{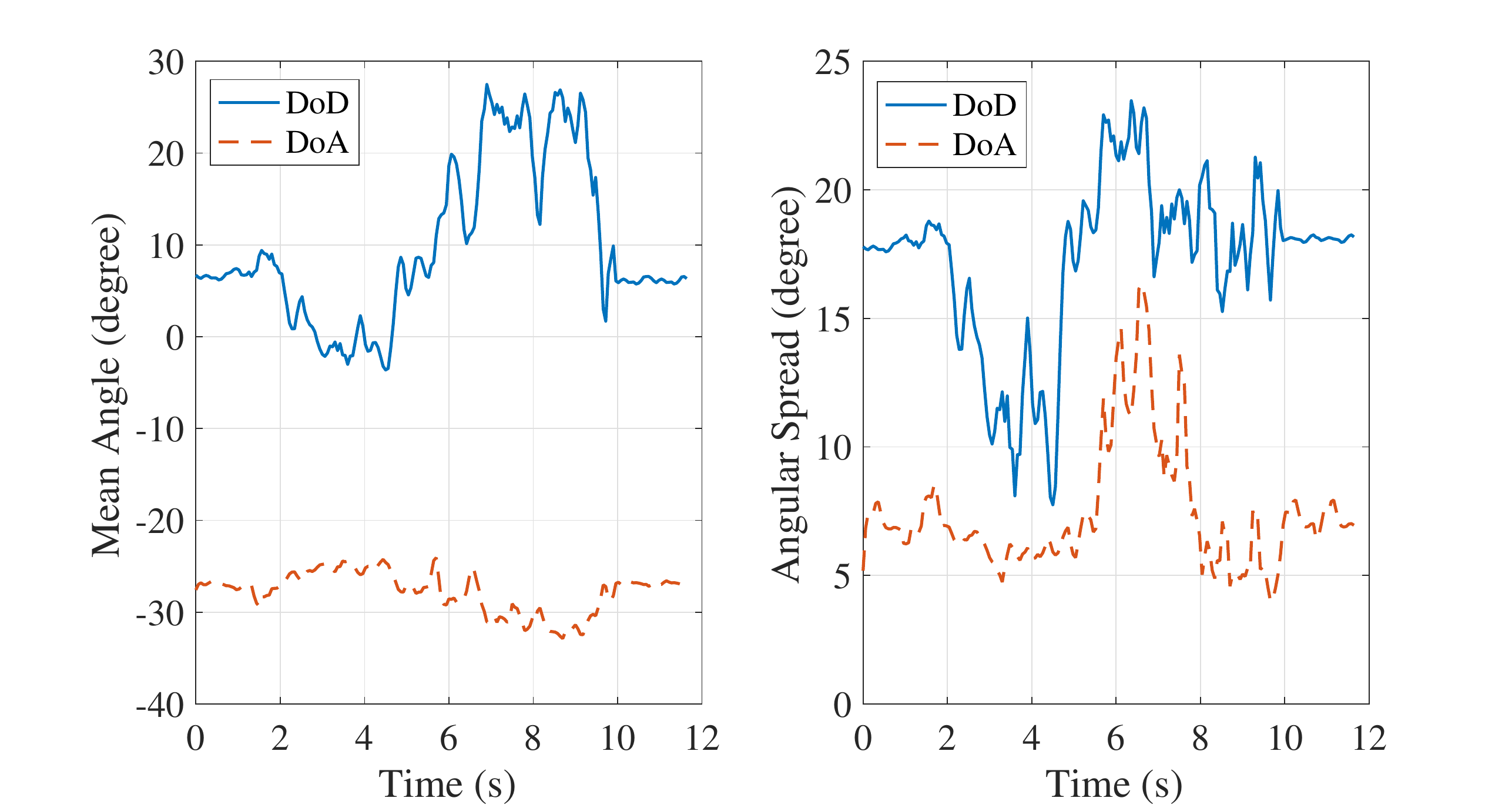}\caption{Mean angle and angular spreads vs time for Case 2}\label{fig:case2_angstats}
\end{figure}

The effects of the truck blocking the paths between TX and RX are visible in the PAS of the TX and RX in Fig. \ref{fig:case2_PAS}. As seen in Fig. \ref{fig:case2_angstats}, due to the two dominant paths with similar DoAs and different DoDs, initially the angular spread for DoA is as small as $7^\circ$ while it is $17^\circ$ for DoD. Once the reflection is blocked (i.e., $t\in[3,4.8]$), the angular spread for the TX decrease to approximately $10^\circ$. Furthermore the TX mean angle also shifts more towards the direction of the LOS path. As the truck clears away from the reflection, for a short duration the PAS and the resulting statistics look similar to the initial case. As the truck blocks the LOS path (i.e., $t\in[5.5,9]$), the mean angles for the TX and RX move toward the reflection which becomes the dominant path in this ``NLOS" channel. Furthermore, the angular spread increases for both the TX and RX during this blockage. Finally, once the truck moves away (i.e., after $t=9$), the channel goes back to the initial idle state.

\begin{figure}
	\centering
	\begin{subfigure}[b]{0.75\linewidth}
	    \includegraphics[width=1\linewidth]{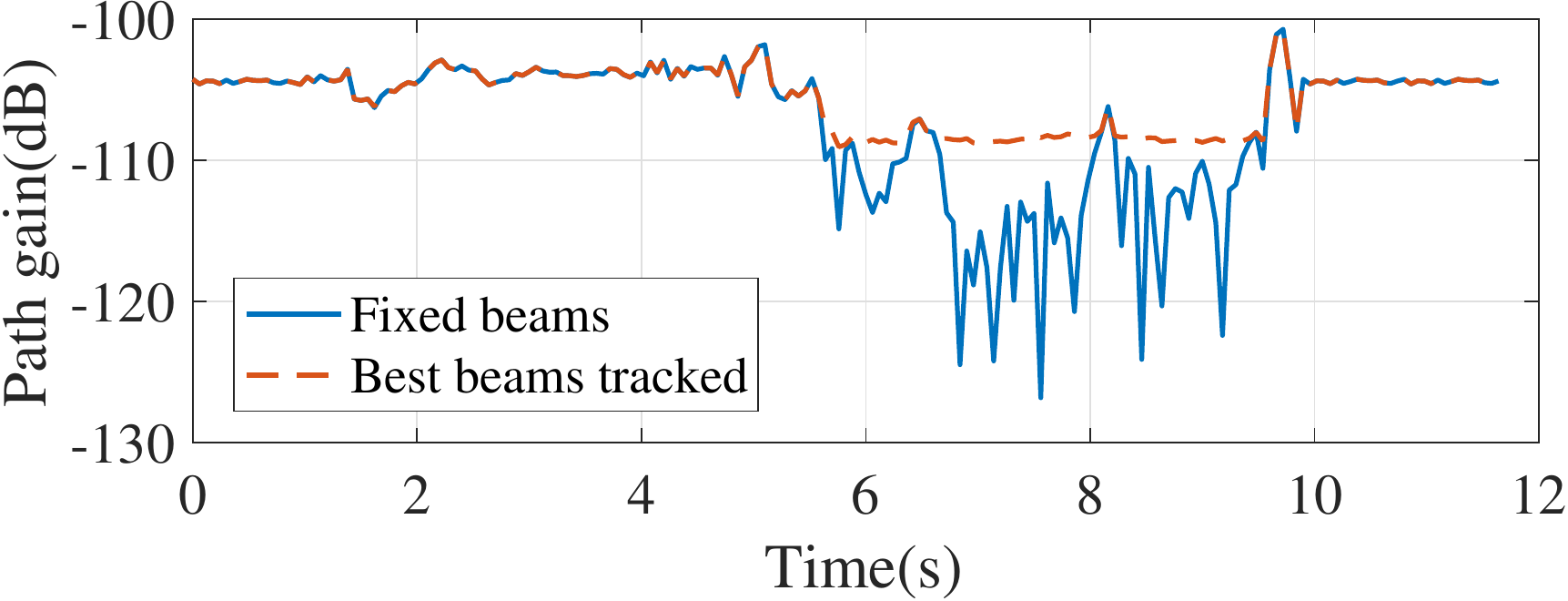}
	    \caption{Truck passes through LOS between 5~s and 9~s}
	    \label{fig:truck}
	\end{subfigure}
	\vspace{0.1in}
	\begin{subfigure}[b]{0.75\linewidth}
		\includegraphics[width=1\linewidth]{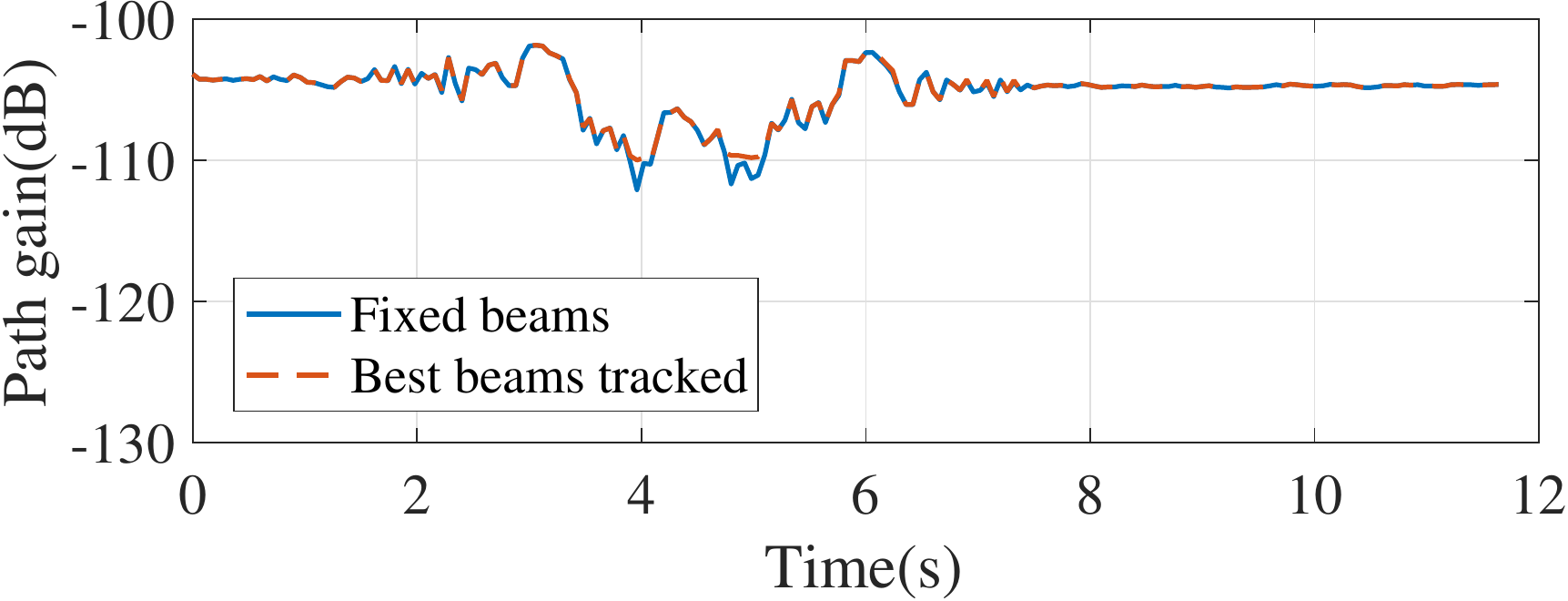}
		\caption{Van passes through LOS between 2~s and 7~s}
		\label{fig:van}
	\end{subfigure}
		\vspace{0.1in}
	\begin{subfigure}[b]{0.75\linewidth}
		\includegraphics[width=1\linewidth]{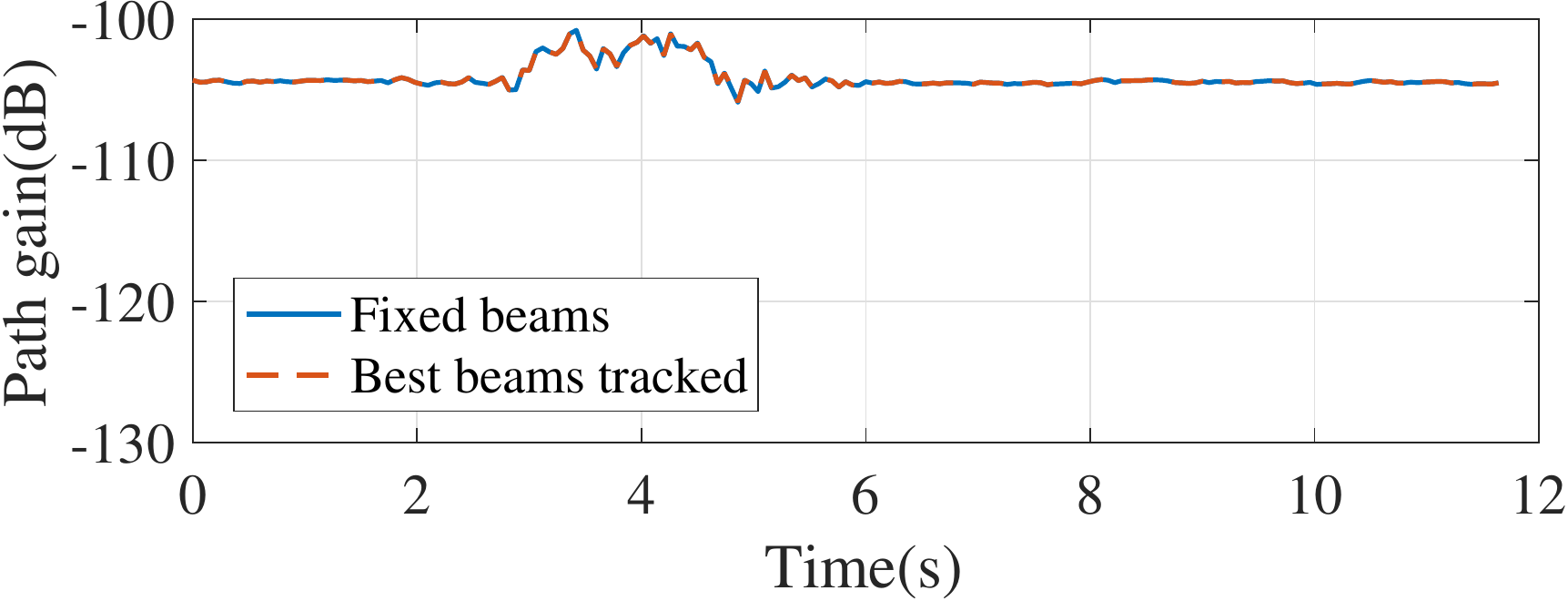}
		\caption{Car passes through LOS between 3~s and 5~s}
		\label{fig:car} 
	\end{subfigure}
\caption{Path gains (dB) vs time: i)RX and TX tracks the best beam pair,  ii) stays on the which best on the average for different vehicles passing through LOS between TX and RX} \label{fig:pass}
\end{figure}

Figure \ref{fig:pass} shows the path gains for the fixed beam with highest average power and if we track the best beam at all times for three different types of vehicles passing between TX and RX. Fig. \ref{fig:truck} is the case with the truck  discussed in the previous paragraph. During the blockage by the truck, it is clear that if TX and RX switch to the beams pointing towards the reflector, the path gain improves by more than 10~dB. Figs. \ref{fig:van} and \ref{fig:car} show evolution of the path gains for similar scenarios for a van and a car. 
In case of the van, although the LOS is blocked, the loss is limited to 6~dB, since it is only a partial blockage. Consequently, tracking the best beam only improves the path gain by 2~dB for a short duration, see Fig. \ref{fig:van}. The car's height is not enough to block the LOS, and its metal roof further introduces another strong reflection. Consequently, during the passage of the car, the LOS gets even stronger and the corresponding beams stay as the best throughout the whole snapshot.

\section{Conclusion}\label{sec_conc}

In this paper, we presented a novel mm-wave channel sounder that can perform double-directional measurements in dynamic environments. By using a beam-forming array, we decreased the measurement time from tens of minutes to milliseconds compared to the rotating horn antenna approach. Furthermore, thanks to the beam-forming gain, the measurable path loss we achieved for a 400~MHz bandwidth is 159~dB without waveform averaging and FFT processing gain. We also validated the sounders phase stability, which is the paramount feature of the proposed design. Compared to the rotating horn antenna channel sounder, which can only measure {\em either} directional {\em or} dynamic channel properties at a given time, the channel sounder presented in this paper can simultaneously estimate DoD, DoA, delay, and Doppler in a dynamic channel. Furthermore, it can provide the temporal variations of the angular spectrum, which is a crucial input for designing beam-forming algorithms. 

\section*{Acknowledgements}

We thank Sundar Aditya, Vinod Kristem, He Zeng, Daoud Burghal and Ming-Chun Lee for their help in the experiments.

\bibliographystyle{IEEEtran}
\bibliography{mmwave}

\end{document}